# Social Centralization and Semantic Collapse: Hyperbolic Embeddings of Networks and Text[1]

Linzhuo Li[a], Lingfei Wu[a,b,c], and James Evans[a,d,1]

[a] Sociology Department and Knowledge Lab, University of Chicago, 5735 South Ellis Avenue, Chicago, IL 60637, USA
[b] School of Computing and Information, University of Pittsburgh, 4200 Fifth Ave, Pittsburgh, PA 15260, USA
[c] School of Journalism and Communication, Nanjing University, 22 Hankou Rd, Nanjing, China, 210008
[d] Santa Fe Institute, 1399 Hyde Park Rd., Santa Fe, NM 87501, USA

Corresponding Author: Email: jevans@uchicago.edu

**Abstract**
Modern advances in transportation and communication technology from airplanes to the internet alongside global expansions of media, migration, and trade have made the modern world more connected than ever before. But what does this bode for the convergence of global culture? Here we explore the relationship between centralization in social networks and contraction or collapse in the diversity of semantic expressions such as ideas, opinions and tastes. We advance formal examination of this relationship by introducing new methods of manifold learning that allow us to map social networks and semantic combinations into comparable hyperbolic spaces. Hyperbolic representations natively represent both hierarchy and diversity within a system. In a Poincaré disk—a two-dimensional hyperbolic embedding—radius from center traces the position of an actor in a social hierarchy or an idea in a semantic hierarchy. Angle of the disk required to inscribe connected actors or ideas captures their diversity. We illustrate this method by examining the relationship between social centralization and semantic diversity within 21st Century physics, empirically demonstrating how dense, centralized collaboration is associated with a reduction in the space of ideas and how these patterns generalize to all modern scholarship and science. We discuss the complex of causes underlying this association, and theorize the dynamic interplay between structural centralization and semantic contraction, arguing that it introduces an essential tension between the supply and demand of difference.

[1] The authors wish to thank Hauchuan Cui and Shahab Asoodeh for helpful comments and discussion. We acknowledge funding support from AFOSR grants FA9550-15-1-0162 and FA9550-19-1-0354, NSF grant 1158803 and 1829366, and DARPA grant HR00111820006.

**Author Biographies**

**Linzhuo Li** is a PhD candidate in the Department of Sociology at the University of Chicago. Before that he earned his BA and MA degree from Tsinghua University. His research examines economic and cultural change, and seeks to understand how order emerges, maintains or breaks down in cultural and economic systems. In addition to this project, he has explored rhetorical and ideological transformation in China's newspaper *People's Daily* from 1946 to 2003.

**Lingfei Wu** is an assistant professor in the School of Computing and Information at the University of Pittsburgh. Before that he earned a PhD in Communication from the City University of Hong Kong. His research interests focus on the future of research, education, and employment in an AI economy driven by knowledge production. His research has been published in peer-reviewed journals, including Nature, PNAS, Scientific Reports, etc., and was covered by the New York Times, Forbes, the Atlantic, and other top media outlets. He is also a core, founding member of the Swarma Club, one of the largest and most active open-science communities in China.

**James Evans** is Professor of Sociology, Director of Knowledge Lab and the Computational Social Science program, and member of the Committee on Conceptual and Historical Studies of Science at the University of Chicago, and External Professor at the Santa Fe Institute. His research focuses on the collective system of thinking and knowing, ranging from the distribution of attention and intuition, the origin of ideas and shared habits of reasoning to processes of agreement, certainty, and understanding. He is especially interested in innovation—how new ideas and technologies emerge—and the role that social and technical institutions like the Internet, markets, and collaborations play in collective cognition and discovery.

**Introduction**

Despite fits of nationalism, temporary tariffs and abortive border restrictions, 'its a small world after all'. Transportation and communication technology from automobiles and airplanes to telephones and the internet have made the modern world more connected than ever before. The march of global media, migration, trade and education across these pathways has gone hand in hand with convergence (Harari 2014) and polarization (Huntington 2000) in global culture. The global human population approaches 8 billion, but human languages are going extinct at an accelerating pace. Nearly 50% of the world's 7,000 languages are on the cusp of being forgotten (Malone 2005), as global communication becomes centralized around widely used languages like Chinese, Hindi, English, Spanish and Arabic, which reduce the space of ideas that can be natively imagined and shared (Whorf and Chase 1956; Regier 2008). Music videos from Luis Fonsi's "Despasito" to Psy's "Gangnam Style" are watched billions of times and popular music charts by country manifest little variation, suggesting a global alignment of musical tastes around centralized styles. Similar trends are seen in the consumption of other cultural products. More Nigerians are watching American movies than ever before, and more Ghanans are watching Nigerian films. More books are shared, borrowed and read across more libraries than ever before, leading to a growing global conversation about a collapsing set of ideas.[2]

The same patterns are manifest at individual scales. Classic studies of group conformity have shown that low status social actors often mimic the opinions and behaviors of social elites in order to fit in, shrinking the space of ideas, opinions and expressions (Asch 1956; Danescu-Niculescu-Mizil et al. 2013). Moreover, 21st Century computational studies of communication document linguistic copying to ingratiate and persuade (Danescu-Niculescu-Mizil et al. 2012; McFarland, Jurafsky, and Rawlings 2013), just as sociologists demonstrate how shared cultural tastes influence the selection of new network ties (McPherson, Smith-Lovin, and Cook 2001; Lewis and Kaufman 2018). Birds of a feather flock together, just as new social ties channel, correlate and collapse the population of cultural tastes (Rogers 2010; Rogers and Shoemaker 1971; Valente 1996).

---

[2] Based on data from OCLC's WorldCat data up until 2008 across 82,000 global libraries with share catalogs.

In this paper, we do not attempt to untie the causal knot of society and culture. Increased social connectivity forces semantic forms like language, music and ideas to compete for collective attention, just as increased linguistic and semantic convergence accelerates connection by reducing communicative friction. Instead, here we seek to acknowledge and formalize the deep connection between social centralization and semantic convergence, building on recent attention to the complex duality of social structure and culture (Mohr 2000; Lee and Martin 2018; Breiger and Puetz 2015; Basov and Brennecke 2017).

We first review recent studies that highlight the relationship between social centralization and semantic diversity. We then introduce a new geometric approach to data analysis that allows us to systematically examine the association between network centralization and the semantic diversity it can sustain in the case of 21st Century physics. We map networks of collaborative relationships and co-occurring concepts through manifold learning into comparable, two-dimensional hyperbolic spaces or Poincaré disks. A Poincaré disk can naturally and simultaneously render both hierarchy in networks and diversity in expressions. We illustrate the metrics defined on these spaces with an empirical exploration of shifts in the centralization of collaboration and those in the diversity of topics and ideas investigated. We show how increased density and centralization in coauthorship ties is associated with sharp reductions in the diversity of ideas. Next, we show how these patterns generalize to all modern science and scholarship. We conclude by theorizing the dynamic interplay between social expansion and semantic contraction and consider its implications for multicultural societies and the maintenance of diversity.

**Centralized Networks and Semantic Diversity**

Recent research has experimentally and observationally identified a tight relationship between network structure and idea diversity (Carley 1986, 1991; DiMaggio 1987; Erickson 1988; Kandel 1978; Krackhardt and Kilduff 1990; Krohn 1986; Mark 1998). Research on the wisdom of crowds has shown that crowds appear wiser than individual participants in stock markets, political elections, and quiz shows (Surowiecki 2004), but social centralization undermines crowd wisdom by reducing independence (Lorenz et al. 2011). One experiment involving a complex estimation task demonstrated that centralized networks tended to decrease the diversity of crowd opinions away from the truth, while decentralized networks decreased diversity toward it (Becker, Brackbill, and Centola 2017). Scholars have also examined these patterns naturalistically in the epistemic culture of biomedicine (Cetina 2009). Our own recent research extracted claims about drug-gene interactions from thousands of papers in the biomedical literature, then aligned them with a massive, robotically-assisted "high-throughput" experiment[3] to evaluate the likelihood of replication (Danchev, Rzhetsky, and Evans 2019). Centralized communities were much more likely to publish new scientific claims that agreed with their own prior work, but were less likely to replicate.

This relationship between social centralization and diversity collapse was further explored in a field experiment where 733 programmers competed in an algorithm development task under two alternative communication and disclosure regimes (Boudreau and Lakhani 2015). In the "open" regime, programmers shared intermediate code products and converged rapidly to the most successful recent advances. Under the closed or competitive regime, programmers competed without communication and a much more diverse range of approaches were cultivated.

This negative association between network centralization and system-level diversity may appear at odds with network analyses that demonstrate how the formation or activation of bridging ties in a network, which typically increase overall network centralization, are also associated with increases in the diversity, creativity and ultimate performance of the person or organization possessing them. In Granovetter's classic article "The Strength of Weak Ties" (1973) and his subsequent book, *Finding a Job* (1974), he demonstrated that the renewed activation of weak-ties enabled flows of novel employment information. Similarly, Burt has shown that managers with ties that bridge distinct social clusters within a company tend to generate broader and better strategic ideas (2004), and Stark and Vedres have shown that overlapping entrepreneurial ventures reap performance benefits (2010). Connections, especially when they bridge cognitively diverse groups and design experiences, can be valuable for the generation of creative products ranging from hit video games (de Vaan, Vedres, and Stark 2015) to research articles (Uzzi et al. 2013).[4]

---

[3] NIH-funded LINCS L1000 experiment, which tested approximately 1.4 million drug-gene associations.
[4] Research on cities often focuses on not only the combination of difference, but its creation through specialization, but the equation between cultural difference generation and erosion is never explicated.

What remains unevaluated in these studies is how the increased internal diversity for those with bridging ties influence future system-wide creativity. Studies have shown that the information benefits from bridging structural divides would erode if everyone pursued a bridging strategy (Buskens and van de Rijt 2008). Nevertheless, the perceived ubiquity of social and cultural barriers has prompted many to advocate greater interdisciplinarity and boundary-crossing creativity (Bauer 1990; Braun and Schubert 2003) without considering its cost for the cultural diversity that might limit returns from future expeditions. Most work on connections and creativity focuses on current, individual-level advantages, rather than counting costs for the entire system over longer time scales.

Scholars have begun to consider the limits of systemic creativity that arise from continuous or repeated connections (Burt 2007). Studies of creative teams reveal the diminishing marginal creativity of repeated collaborations in scientific and scholarly research, product innovation (Guimerà et al. 2005), and Broadway musical creation (Uzzi and Spiro 2005). March's simulation of organizational learning styles takes this to the global level by exploring the consequences of centralized organizational communication. Through simulation, he shows that when people learn quickly from central organizational dogma, diversity collapses and the organization cannot incorporate novel knowledge of its members because through rapid conformity they have forgotten it (1991). This suggests a mutually constitutive relationship between individual-level innovation (Burt 2004) and system-level conformity (March 1991) associated with bridging and centralizing network connections. Centralizing and connecting disconnected societies may enable semantic mixing in the short-term, but through competition and selective cultural extinction, future periods have fewer meanings to mix. The more distant the groups connected, the greater the immediate innovation potential for the socio-semantic "arbitrageurs" that link them, but the larger decrease in system-level diversity that results from putting more semantic possibilities in contact and competition. This suggests that the second law of thermodynamics, where systems spontaneously drift toward greater disorganization or entropy over time, may have a manifestation in the realm of semantics.[5] Meanings in a social system become more *evenly distributed* as social relationships

---

[5] While the Second Law of Thermodynamics assumes the system moving toward complete entropy is isolated, even dissipative systems that exchange energy with their environments manifest increased entropy production over time up to some limited entropy

centralize, increasing the common exchange, competition and collective shrinkage of information (see Appendix and Figure A1).

The correlation between social connection and cultural collapse may not manifest in all contexts. Indeed, it is the centerpiece of classical economics that exchange incentivizes diversity: if I specialize to create more of this and you produce more of that, then through exchange we both get more of what we want. This has been heralded as the principle behind successful "long-tail" Internet retailers like Amazon, who market themselves as able to sell small numbers of vast varieties of products (Anderson 2006). This could suggest that in the hyperconnected Internet age, people are spreading out rather than concentrating their choices. Detailed research on who buys what, however, demonstrates that the long tail of product variety is extremely thin, with more people than ever before spending more of their money on popular products at the head of the distribution (Elberse 2008) and simultaneously purchasing novelties in the tail (Goel et al. 2010). As in product markets, it has also been observed that greater urban density is associated with greater innovation and diversity (Carlino, Chatterjee, and Hunt 2006; Bettencourt et al. 2007), but this fails to account for potential losses in the diversity of rural and other urban areas feeding cities. In short, while it is unclear whether greater social connection always correlates with cultural convergence, empirical examination of potential counter-examples including long-tail product markets and city diversity do not unequivocally argue against it. Moreover, cause and effect among socio-semantic dynamics are also complex and disentangling it is beyond the scope of this article (but see the Appendix and Figure A10).

In this paper we seek to provide a new theoretical articulation and empirical illustration of the widespread relationship between centralization within a system and the diversity that system may sustain. In order to reformulate our thesis about social centralization and semantic collapse to facilitate its direct empirical assessment, we now explore geometric social and semantic representations that build on social-semantic network analyses, which were first to simultaneously render social actors and semantic expressions (Roth and Cointet 2010; Saint-Charles and Mongeau 2009).

---

(Ilya Prigogine and Defay 1944; I. Prigogine 1947). This principle has been widely used in explaining chemical, biological and even social phenomena.

**The Geometry of Social and Semantic Networks**

Why would we want to represent social and semantic networks in a multi-dimensional geometric space? What justifies the complexity of introducing a new paradigm for representing network data? Network scientists historically considered only the topology of social networks, and not their intrinsic geometry. The oldest of us recalls being taught in graduate school that embedding a network in geometry would not help to reduce its complexity because an *n*-node network required *n*-1 Euclidean dimensions to represent. This had actually been disproven in 1984, when the Johnson-Lindenstrauss lemma (1984) demonstrated that networks can be represented in far fewer (≤log *n*) dimensions. The rise of autoencoders, or neural network models that embed data from one representation into another, alongside improvements in high performance computing have recently enabled analysts to efficiently and practically describe social networks with a *much lower dimensional* geometric representation (Perozzi, Al-Rfou, and Skiena 2014; Shuicheng Yan et al. 2005).

Language and meanings, by contrast, have long been represented within continuous geometric spaces and only more recently tortured into network form. Quasi-continuous meaning spaces have been theorized for a century (Wittgenstein, Anscombe, and Wittgenstein 1953; FIRTH and R 1957), and first empirically performed with semantic surveys and factor analysis in the 1950s (Osgood, Suci, and Tannenbaum 1964). Later approaches used matrix factorization on modest text data (Furnas et al. 1988), and more recently auto-encoders like Word2Vec (Mikolov et al. 2013), GLoVe (Pennington, Socher, and Manning 2014) and BERT (Devlin et al. 2018) on very large-scale text. The resulting word embeddings reveal semantic compositionality (Bolukbasi et al. 2016) and recover widespread and subtle cultural associations (Caliskan, Bryson, and Narayanan 2017; Garg et al. 2018; Kozlowski, Taddy, and Evans 2018).

Embedding social and semantic networks into the same low dimensional geometric representation could allow us to directly measure and compare their complex properties. Specifically, we will show that by projecting social and semantic networks onto a hyperbolic space, social centralization and semantic diversity become trivial properties of this new space (Papadopoulos et al. 2012; Krioukov et al. 2010; Nickel and Kiela 2017; Chamberlain, Clough, and Deisenroth 2017).

**The Hyperbolic Geometry of Social and Semantic Networks**. What is hyperbolic geometry? Euclidean or straight-line geometry was constructed to obey Euclid's parallel postulate, which states that within a two-dimensional plane, for any given line ℓ and point *p* not on ℓ, there exists exactly one line through *p* that does not intersect ℓ. This constructs a Euclidean space characterized by "straight" vectors and "flat" planes. Hyperbolic geometry is not straight, but curved, violating the parallel postulate such that (infinitely) many lines may go through *p* without intersecting ℓ. A triangle has 180 degrees in flat Euclidean space, but fewer in hyperbolic space. A mountain pass, coral reef, or crocheted frill furnish examples of negatively curved surfaces that can be rendered with just two hyperbolic dimensions. Figure 3, panel A presents M.C. Escher's print *Circle Limit IV* and panel B shares the underlying Poincaré disk model, which projects a 2-dimensional hyperbolic space onto a 2-dimensional Euclidean space for visualization (cited and recreated from Dunham and Others 2009). Escher's print alternatively tessellates angels and devils on the disk, *ad infinitum*, revealing how the space produces a "fisheye view" that packs both a single focus and wide context into the same frame. In fact, the image details become exponentially more dense as the radius increases.

———————————————

Figure 1 about here

———————————————

Why model social and semantic networks with a hyperbolic space? Social networks are defined by clusters (Watts and Strogatz 1998; Freeman 1977) bridged by network brokers (M. S. Granovetter 1973; Burt 2004, 2009). These two properties are captured in the "small-world" model of Watts and Strogatz (1998). Social networks also manifest diversity in number of connections such that a few social actors possess most connections, while most possess a few. This results in phenomena such as the friendship paradox, where most people's friends have more friends than themselves (Zuckerman and Jost 2001; Feld 1991) and has been modeled by a process in which network agents prefer connection to those already popular, resulting in "scale-free" distributions such that any agent in the network has the same relative proportion of those with immediately more and less connections (Simon 1955; Barabasi and Albert 1999).[6]

Physicists Papadopoulos and colleagues showed how both "small-world" and "scale-free" properties can be reproduced by translating the process of network linkage into a hyperbolic space such that social actors connect to others within a fixed radius (Papadopoulos et al. 2012; Krioukov et al. 2010). This approach uses the center-periphery asymmetry of hyperbolic space to model the global reach and popularity of central social actors. Actors at the center of a hyperbolic space can reach all other nodes while those at the periphery are exponentially distant from one another and must go through those at center to efficiently reach others (see Figure 1). A social network embedded in hyperbolic space is characterized by two directions, the "popularity" direction pointing from center to periphery, which captures the network's centralization, and "similarity" along the periphery where nodes are locally clustered, which captures the network's diversity. This interpretation is supported by Chamberlain and colleagues who embedded social networks into hyperbolic space and found that actors with high betweenness (Freeman 1977) were central in that space, whereas high degree nodes with low betweenness were pushed to the periphery (2017).

Hyperbolic space is also a powerful representation for human languages, where bridging conjunctions like "and" and "but", common verbs like "to be" and "to have", or words with multiple meanings like "book" or "bear", may connect otherwise disconnected words and clusters of meaning. This builds hierarchy in semantics comparable to social networks (Fellbaum 2005). Hyperbolic geometry is a strong candidate for modeling these bridged, asymmetric structures. Because of the expanded area or "ruffled edge" of hyperbolic space, two concepts can be simultaneously close to the same central, bridging idea but far from one another. As a result, hyperbolic geometry captures the hierarchy common to semantic systems. For example, WordNet is a network of words connected by shared meanings of synonymy. Embedding WordNet in hyperbolic space requires only a few dimensions (2-5) to model the complex relationships between words, far less than the hundreds of dimensions required to model it in Euclidean space (Fellbaum 2005; Chamberlain, Clough, and Deisenroth 2017; Nickel and Kiela 2017). Like a social networks, a semantic network embedded in hyperbolic space is characterized by two directions, the "frequency" direction pointing from most to least common words, which captures centralization in the semantic space, and "similarity" along the periphery where meanings are locally clustered, which captures diversity among those meanings.

---

[6] Specifically, the "scale-free" property imples that when scaled by the correct exponent $\tau$, the logged distribution of social connections can be represented as a straight line. For example, if $\tau = -1$, then if one person had a million Twitter followers, a million people would have one follower.

Geometric embedding models also operationalize and extend key elements of Bourdieu's field theory. Bourdieusian cultural fields offer a model for how individuals, cultural objects, and positions in social structure are located relative to one another in structurally homologous "social spaces," with relations between entities described in terms of "distances" (Bourdieu 1989; Kozlowski, Taddy, and Evans 2019). Bourdieu represented these spaces geometrically using correspondence analysis, representing distances between entities and meaningful dimensions through placement in a dimensionalized Euclidean plane (Bourdieu 1984). The qualities of social centralization and semantic diversity, however, manifest structural properties of hierarchy rather than dimension. Recent research demonstrates a powerful identity between social hierarchy in productive communities and organizations, on the one hand, and hierarchy and diversity in knowledge, whereby social structures facilitate the more or less efficient matching of problems to problem solvers (Garicano 2000). Hyperbolic embeddings are precisely tuned to capture this hierarchy characteristic of social and semantic organization.

But what justifies the complication of embedding social and semantic networks to hyperbolic space? We are about to analyze the case of academic physics, which presents its own illustrative example. Isaac Newton represented the universe in simple, three dimensional Euclidean space, but the process of gravitational attraction between bodies in that space was complex—the product of the bodies' masses and the universal gravitational constant divided by the square of their distance. Albert Einstein represented the universe in a more complex, warped space-time, but the process of gravitational attraction was simply the distance between objects in that space. A geometric (and hyperbolic) representation of social and semantic networks may seem like a complicated new addition to the marketplace of cultural analytics. But by efficiently packing the complicated asymmetries of those networks into curved hyperbolic space, the focal properties of centralization and diversity become trivial distance measurements. We embed social and semantic relationships within comparable hyperbolic spaces to directly interrogate the relationship between patterns of social centralization and semantic diversity.

We empirically examine social and semantic entanglement in the context of scientific knowledge production. Many studies have demonstrated that social influence contributes to both inequality and unpredictability in cultural markets (Salganik and Watts 2008; Strang and Macy 2001), yet there has been less empirical research on whether or how scientific understanding is driven by the consensus of scientific communities (Kuhn 1970). As highlighted by Hutchinson (2011), scientism—the view that science has its own logic and the objective natural science is the only legitimate origin of knowledge—remains popular, but provokes culture wars. By showing the entwined social and semantic influences on scientific investigations underscores the social embeddedness of science as a cultural industry.

Specifically, we analyze the link between centralization and diversity across science and scholarship, with a special investigation of the semantic domain of academic physics, as represented by the complete 21$^{st}$ Century publications of the American Physical Society (APS), a society organized at Columbia University in 1899 "to advance and diffuse the knowledge of physics". Insofar as the hyperbolic embedding of socio-semantic structure represents a kind of *physics of sociology*, we felt it might be appropriate to recursively use it to perform a *sociology of physics*. We specifically examine patterns of collaboration between universities in terms of their publishing physicists, and the diversity of physics topics those universities publish. We use published physics rather than a more ephemeral socio-semantic context like chit-chat on Tumblr, Twitter or Facebook largely because it is much more difficult to identify individual persons and trace them persistently in social media settings, even though it may be easier to follow micro-social and semantic interactions in those settings. Inflamed interaction between a flock of tweets is more specific and intense than that between an archive of articles referencing one another. Because academic physicists have a strong incentive to build persistent reputations through publishing within a single area, our data represents a very conservative case from which to examine the relationship between of social centralization and semantic diversity. If we observe that collaboration and topic selection move together among academic physicists, it is likely amplified elsewhere.

We note that while most prior work has focused on collaborations between individuals rather than institutions, the number of ties across institutions closely tracks the total number of ties overall (see Figure 2). By focusing on coauthorships among institutions, we are able to see how ties that bridge institutional collaboration clusters relate to reductions in topic diversity across the entire system. Moreover, centralization within fields is often discussed in terms of the stable rank order of dominant institutions that integrate the broader field.

Formally, when we embed networks to a Poincaré disk (2D hyperbolic space), the position of each node $i$ is defined by two parameters in polar coordinates that capture these properties, radius $r_i$ and angle $\theta_i$. The radius quantifies position in the hierarchy: nodes of small radius literally hold a central position. The angle between two nodes $\theta_i - \theta_j$ quantifies their proximity or structural similarity, and the entropy over all such angles is a direct measure of diversity.

**Data**

We analyze centralization and diversity in science and scholarship using physics articles from the American Physical Society (APS) journals, and then confirm some of these findings with articles archived in the Web of Science (WoS). Because the identification of co-authorship and institutional networks hinges on how author and institution names are disambiguated, we discuss our approach to disambiguation as we introduce our datasets. First, we introduce the WoS data on which our disambiguation approach hinges, and then the APS data that draws on it.

---

Figure 2 about here

---

Our WoS dataset contains 43 million journal papers and 615 million citations that span from 1900 to 2014. These papers are published across 15,146 journals. To match the APS data (introduced below), we selected and analyzed 13 million journal articles from our WoS dataset across seven fields during 1994-2013 based on WoS journal classification meta data.

After performing name disambiguation on this data (see Appendix for details), we connected name-disambiguated institutions to name-disambiguated authors in each year, enabling authors to change institutions across years. In total, we obtain 372,495 authors from 28,754 institutions. The most institutionally connected author has 26 institutional affiliations, and 83% have only a single one.

We weighted author collaboration networks by iterating over all papers for a given year and connecting scholars that collaborated on the same paper. In this way we obtained ten annual networks of author collaboration for 2002 through 2011. Next, we renormalized the networks by merging scholars from the same institutions into single nodes and then aggregated author collaboration edges to connect institutions (see Figure 2). Edges between scholars from the same institution are retained as self-loops. To compare institution networks across years, we fixed the nodes by selecting the top 300 institutions, but removed 10 due to missing data and so we analyzed collaborations between the most prolific 290 institutions. After we constructed the ten yearly institution collaboration networks, we embedded each into a 2-dimensional hyperbolic space using the Poincaré disk model (Nickel and Kiela 2017). As anticipated, we found that the embedding space captured both the hierarchy among institutions and diversity between them, as shown in Figure 3.

In order to study semantic content in physics, we used the Physics and Astronomy Classification Scheme (PACS), developed by the American Institute of Physics (AIP) and used in the Physical Review journals since 1975 to identify fields and subfields of physics. PACS is a hierarchical partitioning of the full spectrum of topics in physics, astronomy, and related sciences. Each PACS code is a six-digit number, in which the first digit indicates one of ten subfields to which it belongs (see Figure 3). PACS codes are semantic artifacts—agreed upon expressions of meaning. We constructed the PACS code co-occurrence network by iterating over all 359,395 papers in our dataset and connecting PACS codes that appeared together in the same paper. The network contained 5,856 nodes, 244,106 edges, with a total edge weight of 1,124,724. After we constructed this PACS code co-occurrence network, we embedded it into hyperbolic space using the Poincaré disk model, comparable with the institutional collaboration embedding described above. As with the organization network, we found that the representations of PACS codes in a 2D hyperbolic embedding captures both their hierarchy and similarity, as shown in Figure 3.

In order to extend our validation of the association between social connection and semantic convergence across many fields of science, we used WoS meta-data from 13 million journal articles and journal classifications[7] 1994-2013 to construct network and information theory measures. We were unable to construct hyperbolic spaces for all of these collaborations and topics, and so instead for each field in a given year, we constructed author collaboration networks and counted the average number of collaborators for authors as the measure of social densification in that field. The degree of semantic convergence is measured by the entropy of word distributions. Specifically, we extracted the abstracts of articles and selected all tokenized words that occur twice or more, then after removing stopwords, calculated the entropy of word distributions for each field in a given year. To control for differences in word vocabulary length across fields, we follow the convention of normalizing calculated entropy by its theoretical maximum, Log(N), for which *N* is vocabulary length. We examine the association year by year at different levels of resolution, and detailed results are shown in Figure A7 for 13 fields, in Figure A8 for more than 200 subfields, and in Figure A9 for approximately 10,000 journals.

**Methods**

We use the Poincaré Embeddings algorithm (https://github.com/facebookresearch/poincare-embeddings) developed by Facebook AI scientists Nickel and Kiela (2017) using PyTorch. This algorithm automatically learns hierarchical representations of nodes in networks by training on positive and negative samples by using a shallow, three-level neural network auto-encoder. We embedded our (1) university collaboration network across publications in all time periods (2002-2011), (2) separately for each year, and (3) the co-occurrence network of all PACS codes across publications in all years, using a 2-dimensional Poincaré disk, with a learning rate of 0.5 and a negative sample size of 50.[8] This produced hyperbolic embeddings in which each node $i$—a university in the collaboration embedding and a PACS code in the semantic embedding—has radius $r_i$ and angle $\theta_i$. Nodes of small radius hold central positions in the circularly arrayed hierarchy. Angles between two nodes, $\theta_i - \theta_j$ quantifies their social or semantic difference (See appendix for the definition of inner product of two nodes in the 2d Poincaré disk).

---

[7] The journal classifications are used to map articles into subfields and fields of science.

In order to calculate the position $\theta_i$ for each university $i$ in the semantic or PACS code space, we selected the peak or mode of the Gaussian kernel density estimation for all of its weighted PACS codes in a given year (see Figure A5) from the PACS code positions estimated for all years, 2002-2011 (see Figure A4). To estimate each university's level in the topical hierarchy, we calculated the hyperbolic distance between all pairs of PACS codes from a university and select the PACS code of the shortest hyperbolic distance to all other codes. This selected PACS code is closest to "the center of mass" in the hyperbolic space and thus represents the diverse topics covered by the university. With this estimation, each university $i$ has a unique $r_i$ and $\theta_i$ inherited from the selected, representative code in both collaboration and topical space for each year.

These preliminaries set us up to directly explore the relationship between social and semantic structure in science and scholarship, and specifically in physics. We first did this by taking each pair of universities and calculating their social and semantic hyperbolic distances at every time point, which generated two time series. We then evaluated these time series pairs in order to explore the complex relationship between social and semantic factors. This suggests that increased institutional contact is associated with a decrease in the diversity of topics explored. We also performed the same analysis between each university and all other universities in the system at every time point to explore the connection between social relationships and the contraction of the entire semantic system.

As mentioned above and detailed in the appendix, we validate our findings with traditional network measures of social density and information theoretic measures of semantic diversity. Nevertheless, we argue that our hyperbolic metrics are conceptually superior. Consider three plausible measures to capture the spread and size of topics in semantic space: 1) the span of vocabulary; 2) that span, weighted by topic frequency—or entropy; and 3) that span, weighted by frequency and accounting for semantic overlapping between words—what our embedding approach accomplishes by representing words as weights on orthogonal dimensions of meaning. Entropy measure can only be an unbiased measure of semantic diversity if all keywords are independent from one another and represent orthogonal dimensions of meaning. We chose hyperbolic embeddings for our measure as they model semantic overlapping and frequencies directly as angles and radius, respectively.

---

[8] We specified a batch size of 30, which limits the size of examples from data on which the algorithm trains. We trained the model on 300 sequential epochs, of which the first 20 were "burn-in" epochs used only to initialize the model. The model was evaluated every 5 epochs, and every time gets updated if it has lower loss against the validation data. The co-occurence of collaboration edges in our network are undirected and so we set the parameter *symmetrize* to be true. In addition, the *epsilon* parameter is set be no smaller than 1e-5 such that the maximum hyperbolic radius is 1-1e-5.

**Findings**

**The Social and Semantic Geometry of Physics.** In order to summarize the global social and semantic structure of 21$^{st}$ Century physics, we calculated and graphed the radius and angle in embedded social space for each of the 290 top physics publishing universities and in semantic space for thousands of PACS codes. We found that in both social and semantic disks, the radius systematically quantifies the institutional "hierarchy" of university physics departments in the inter-university collaboration network, and angles quantify the research diversity between them. As demonstrated in Figure 3C, high-ranking institutions from various global regions occupy the central area in the social space, including Columbia University from North America, Peking University from Asia, and Australian National University from Australia. Geological groups emerge and cluster together on the periphery, including institutions from Canada (University of Waterloo, McMaster University, McGill University, and the University of Alberta), Japan (Kelo University, Waseda University, Osaka City University, and Tokyo Metropolitan University), Israel (Hebrew University and Hebrew University of Jerusalem), and Italy (the Universities of Catania and Radua).

______________________________

Figure 3 about here

______________________________

PACS codes group together on the perhiphy, forming and indicating subfields in physics (Figure 3D). The most clustered PACS codes include "The Physics of Elementary Particles and Fields" (red) and "Nuclear Physics" (orange), which nontrivially overlap in the lower left quadrant of the Poincaré disk. The tall peaks of density distribution centered on these fields suggest that their codes are nearly always used with others from the same field in research articles. They rarely combine with other field codes. Situated at the antipodes of the disk, in the upper right, are "Condensed Matter: Structural, Mechanical and Thermal Properties" (light blue) and "Condensed Matter: Electronic Structure, Electrical, Magnetic, and Optical Properties," (blue) which each have a higher $\theta$ and are substantially more spread. Nuclear physics and the physics of elementary particles focus on individual particles and their properties. Condensed matter physics focuses on interactions between particles, often in the context of large, naturalistic materials. It may not do too much violence to the epistemic cultures of physics to say that particle physics is the *economics* of the physical sciences, flush with cash for large-scale experiments, focused on the essential monads of existence in artificial isolation and furnished with a sense of superiority that their work is more fundamental and important than their neighbors (Fourcade, Ollion, and Algan 2015).[9]

Condensed matter physics might then be viewed as the *sociology* of the physical sciences, fixed on social interactions between particles in complex, overlapping physical fields, often embedded in the rich context of materials and phenomena that can be macroscopically experienced by ordinary humans, like sandpiles (Jaeger and Nagel 1992), coffee rings (Deegan et al. 1997) and the physics of food (Poon 2002). Consider the Ising model from statistical mechanics that sources from "soft matter" concerns and models the emergent property of phase transitions in ferromagnetic materials as atomic spins arranged in a graph. The graph arrangement allows each spin to interact with its neighbors. This model is frequently used by social network analysts and physicists to model human social interaction (Klemm et al. 2003; Wasserman and Pattison 1996). Insofar as physics increasingly tends to cleave between a focus on particles or interactions (see Figure 4C), we leave it to the reader to decide which represents the angel and which the devil in Figure 3A.

---

[9] Wolfgang Pauli denigrated the physics of solids as "Schmutzphysik" or dirt physics, while Murray Gell-Mann jabbed that "solid state physics" be replaced with "squalid state physics" (Montgomery and Largent 2015 p. 195).

With a special hubris, spokespeople in physics have claimed their science as the fundamental center (Young and Freedman 2013) or even the only meaningful science—in the words of Ernest Rutherford: "all science is either physics or stamp collecting." This work highlights that 21st Century physics is not a unified front, but rather a growing polarization between particle and condensed-matter physics, which originates from two fundamentally different perspectives on the decompositionality of our universe (P. W. Anderson 1972). These debates have become cause and consequence of polarized communities and led to real-world dramas, including the cancellation of U.S. Superconducting Super Collider (SSC) in 1993 after $2 billion had been spent, and the 2016 debate and cancellation on the ambitious plan of Circular Electron–Positron Collider (CEPC) in China with an early stage budget of 30-billion-yuan ($4.3-billion). In this way, our investigation into the polarization in physics may provide insight into similar patterns in politics, ideology, and other domains.

Plasma physics, fluid dynamics and classical mechanics lie off of the fundamental particle / soft matter axis, but have also been much less intensively published in 21st Century APS publications.[10] The most spread field is not a field at all, but rather "Interdisciplinary Physics and Related Areas of Science and Technology" (deep blue), which diffusely spreads around the entire system and whose codes are frequently connected to those from other fields. That these codes are not often close to the center of the disk suggests that rather than "Interdisciplinary", they are more likely "Multidisciplinary" codes that interact with other specific fields at the periphery, but do not integrate those fields into the broader system of physics. The field codes closest to the center are those from "General" (red), which is not a category at all, but rather was derived from the first two decimal places of each PACS code and so, by definition, connects different subfields of physics by forming the trunk of their hierarchical tree.

We map changes in the social hierarchy and centralization among university physics collaborations over time, from 2002 until 2011 in Figure 4A, where universities are colored by region. Here we see strong and consistent geographical clustering in collaboration over time, with North American, European, and Asian universities tending to collaborate with one another. Australian and South American Universities are dispersed as satellites of geographically diverse, dominant universities from Asia and the West. Within the hierarchy, North American institutions remain the most central collaborators throughout, with the lowest average radius $r$, but Asian universities experienced the greatest change in $r$, migrating rapidly from the periphery toward the collaboration network's center and resulting in greater centralization in the global collaboration system.

---

[10] Some fields like plasma physics have suffered in the U.S. and Europe from encumbrances for the international collaboration, ITER, the International Thermonuclear Experimental Reactor.

We also chart changes in the semantic position of physics universities over the same time period in Figure 4C, with universities colored by region, as in Figure 3C, but plotted against the subfield locations represented in Figure 3D. It is interesting that the representative positions of most university physics departments cluster near the subfield of "Condensed Matter Physics" (45 degree) or "The Physics of Elementary Particles and Fields" (225 degree). Moreover, the field of university physics departments tends to collapse toward polarization between the physics of particles and of complex materials over time. In 2002, a scattering of European and a few U.S. universities have a wider mixture of PACS codes, focusing on plasma or fluid physics, but by 2011, they tend to represent some mixture of particle or systems physics. North American Universities appear most likely to specialize in particle or systems physics, while European and Asian Universities represent a much more even mix of the two polar branches.

**Social centralization and semantic contraction.** When we compare the social and semantic hyperbolic distances between every pair of universities over all years, binned, we see a very strong positive association, as graphed in Figure 4B. This suggests that social and semantic distances are tightly bound together across all years. We also see that the range of distances is much smaller in the social than the semantic world. Nevertheless, the range of the two distances are very different. Topical distances vary between .4 and 1.4, while social distances move from .5 to 8. This difference reflects the deep topical inertia in academic physics, while patterns of collaboration can shift dramatically from year to year. We also see that the semantic space of university physics is dramatically contracting over time, nearly halving for the least connected institutions between 2002 and 2011.

______________________________

Figure 4 about here

______________________________

Next, we explore the association of collaboration and focus over time. For each pair of institutions, we estimated an OLS regression to identify the slope representing change in social distance over time. The coefficient for time, $\beta_{sij}$, should be interpreted as the speed of social convergence or divergence between any two institutions $i$ and $j$ across the 2002-2011 time series. Then we separately estimated the slope of semantic change over time, where the coefficient on time, $\beta_{cij}$, reflects the speed of semantic convergence or divergence between that pair of institutions. The Pearson correlation coefficient between the two diverging speeds, $\beta_{sij}$ and $\beta_{cij}$, is 0.98 (P-value < 0.001), providing strong evidence that in the context of academic physics, collaboration is associated with topical convergence (see Figure 4D). In summary, not only are university physics collaborations moving with their evolving interests and expertise, but they are moving the same way over time.

In appendix Figure 5, we validate the relationship between social centralization and semantic contraction with standard network and information theory measures. In the inset of the right panel, we see that when an institution's clustering coefficient in the network goes up, their entropy over PACS codes goes down. This suggests with greater density comes less (individual) diversity. We note that the hyperbolic measures between pairs of universities and between each university and all others provides more dispositive and descriptive information regarding social centralization and contraction. The intellectual burden of projecting social and semantic networks to a geometric representation that may be unfamiliar to cultural analysts is directness of the measures of social centralization and semantic diversity that result.

———————————————

Figure 5 about here

———————————————

Finally, we sought to explore the association between centralized collaboration and the convergence of topics beyond physics, but disambiguating authors and assembling or constructing a terminology for all fields was beyond the scope of this investigation. We relied on the established relationship between enlarging teams and social centralization across science (Danchev, Rzhetsky, and Evans 2019; Wuchty, Jones, and Uzzi 2007). We selected and analyzed 13 million journal articles from our WoS dataset across seven fields between 1994 and 2013. Based on the WoS data that codes journals by field, the average number of coauthors increases over time for all fields, which corresponds to the pattern we find for physics (APS) as shown in the inset. This corresponds to a decreasing spread or entropy over the words used in article abstracts for those same fields, which also corresponds to decreased entropy among PACS codes across APS articles. Insofar as greater collaboration is a critical component of greater centralization in APS, we see that greater collaborative centralization in the form of enlarging and overlapping teams is linked with a reduction in the distribution of topics discussed and investigated.

While the trend of decreasing topic spread and increasing collaboration are observed for both physics and beyond overtime, we also found that similar patterns exist cross-sectionally for almost all science disciplines in all years investigated. In appendix Figure A7 we show that for the 13 major fields in science, topic spread is lower for those fields with greater author collaboration. OLS regressions show that slopes for different years are stable between -0.01 and -0.03 and significant for most years. The associations remain consistent at higher resolution for more than 200 subfields (Figure A8) and more than 10,000 journals (Figure A9). These findings add further evidence that social connections and semantic convergence are closely associated, insensitive to which method we use.

**Discussion**

We live in an age of social hyper-connectivity and centralization. From the rise of social media to the drop in transportation costs that facilitate social gathering, meeting, and mixing, there are more opportunities to connect in a growing and increasingly centralized hub of relationships than ever before. In the sociological and organizational literature, the process of bridging social (Burt 2009) and cultural holes (Pachucki and Breiger 2010; Lizardo 2014) has been linked with a wide range of individual, organizational and city-level benefits through increased access to cultural information, and opportunities for strategic and creative recombination. Individual benefits incentivize strategic networking, enabling entrepreneurs to engage in arbitrage across communities and markets (Burt 2009). Scientists and engineers similarly strive to place their ideas and inventions at the center of complex networks in order to cement their continuing influence and importance (Latour 1987). All of these forces reduce the diversity of ideas, symbols and other semantic forms, ultimately placing them in competition and resulting in rising rates of cultural extinction. With growing attention to the duality of social and cultural structures (Mohr 2000; Lee and Martin 2018; Breiger and Puetz 2015; Basov and Brennecke 2017), we argue for the importance of attending to this cycle of social exploration and cultural exploitation that characterizes processes of colonialism, globalization, and capitalism, resulting in cultural creativity, convergence, and sometimes collapse. The process is manifest in the extinction of indigenous languages, practices and artifacts. Moreover, the causal arrow goes in both directions. Shared global languages, opinions and tastes facilitate more social connection and centralization, which recursively put more semantic forms in competition and at risk of abandonment.

In this paper, we demonstrated that social and semantic worlds could be informatively represented with hyperbolic geometry to simultaneously reveal the complex relationship between social centralization and semantic convergence. We argue that the burden of learning this novel representation is justified by the importance of existential and practical benefits for understanding and preserving cultural diversity, which modernity threatens. Hyperbolic embeddings reveal centralization and diversity as simple and direct measurement of the space in which social and semantic networks are embedded.

We then rendered the university collaboration network of $21^{st}$ Century physics traced by the corpus of the American Physical Society within a two dimensional Poincaré disk. We comparably represented the semantic field of physics. We found that social connection among academic physics departments is becoming more centralized and dense over time, and that universities tend to increasingly sponsor research polarized between particle and nuclear physics, on the one hand, and condensed matter and plasma physics, on the other. We ran the same analyses for each university in their collaborations with all other universities, finding that greater external collaboration is consistently associated with future convergence to the most popular topics in the broader field. These findings were supported by a standard analysis of network density and information theoretic diversity. We also demonstrated that increased and enlarged teaming across all areas of science and scholarship, a driving force of collaborative centralization, is associated with a similar drop in the diversity of words used within those fields.

There are several limitations of our analysis. First, our case of modern physics may not be perceived as representing the fluidity and spontaneity of other semantic domains. Nevertheless, we argue that it represents a conservative case, given the strong rewards for topical consistency in academic promotion. Second, our short time frame, 2002-2011, does not allow us to observe long-term change and influence between social and semantic spheres. Third, by calculating statistics like percentages over many estimated models (see Figure A10), we implicitly avoid taking into account the social and semantic dependencies between cases. Hyperbolic distances between university pairs in the collaboration and PACS code networks were constructed in context of all other collaborations and published PACS codes and so their measurements are naturally dependent upon one another. Despite these limitations and complexities, we believe that our illustration demonstrates many ways in which hyperbolic embeddings can make visible and shed light on the complex association between social centralization and semantic diversity so characteristic of modernity.

The great irony is that modern tastes for diverse music, food, clothing and ideas may themselves limit the sustainability of diversity. The tight relationship between social centralization and semantic contraction poses a paradox for modern, multicultural societies that value diversity. Extensive research in social, behavioral and organizational science documents the largely positive effect social and cultural diversity exerts on the collective production of information, goods and services (Joshi and Roh 2009; Page 2008). Individuals from culturally distinct groups embody diverse cultural perspectives and cognitive resources that combine to produce solutions and designs that outperform those from homogeneous groups (Mannix and Neale 2005; Woolley et al. 2010; Hong and Page 2004; Nielsen 2012). Collaborations between inventors from distinct social groups result in more creative patents (Fleming, Mingo, and Chen 2007), scientific teams representing distinct disciplines produce more highly cited papers (Wuchty, Jones, and Uzzi 2007), gender diversity broadens the questions scientists ask (Nielsen 2012), and political diversity leads to improved political pages on Wikipedia (Shi et al. 2017). Beyond performance, some enjoy a taste for difference or heterophily (Lazarsfeld, Merton, and Others 1954). But if and when semantic differences fuse through social centralization, they become less different. And when differences erode, so do the performance benefits that come from them. Social centralization consumes semantic diversity.

Only by considering the relationship between society and meaning on both short and long time scales can we clearly see the essential tension between the enjoyment and protection of semantic diversity. Semantic differences can be generated through isolation and specialization, but as with indigenous languages like Wichita, Mandan and Yurok, or megafauna like the Tasmanian tiger, Chinese Baiji dolphin, or North American heath hen, all of which irreversibly disappeared within the last hundred years, it is easier to extinguish semantic and biological forms than to evolve new ones. We highlight the power of hyperbolic embedding in making diversity and centralization visible, and argue for greater incorporation of long-run cultural dynamics into theories of social centralization, integration and change to insure long-run benefits from sustainable diversity.

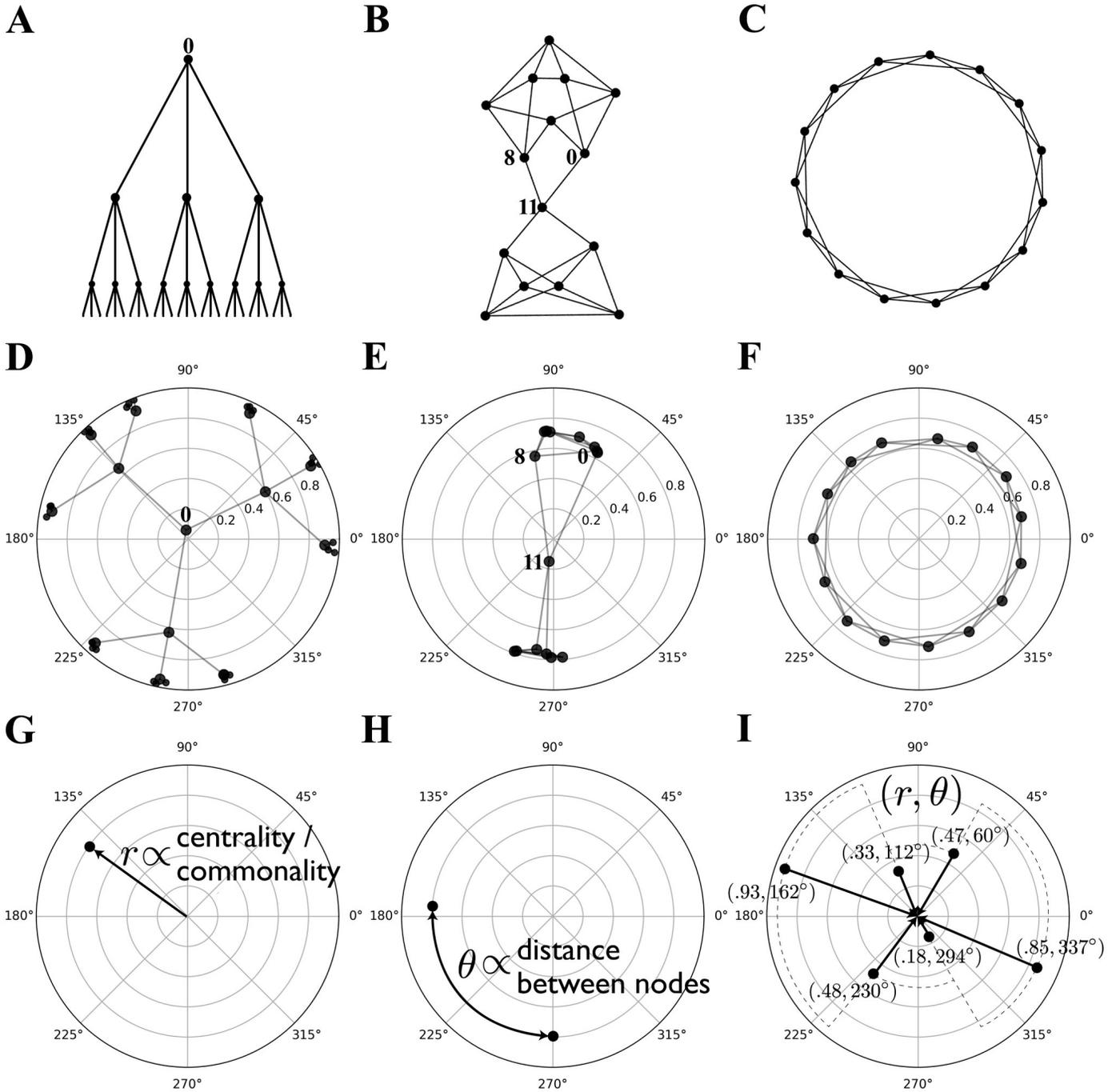

**Figure 1.** A simplified illustration of hyperbolic embedding. **A**, **B**, and **C** present three graphs designed to have different hierarchies (**A** > **B** > **C**). **A** is a tree of branching factor three and across three levels. **C** is a ring lattice in which each node connect to its four nearest neighbors. **B** is created by randomly rewiring edges in **C** while keeping node degrees unchanged and the whole network connected. Specifically, to obtain hierarchical structures we run the rewiring 200 times and update the graph only when (a) it remains fully connected and (b) the standard deviation of betweenness centralities for all nodes in the graph increases. **D**, **E**, and **F** show the 2D hyperbolic embedding in Poincaré disks of graphs **A**, **B**, and **C**, respectively. **G**, **H**, and **I** describe the embedding of hyperbolic space. **G** shows how the radius of the node $r$—its distance from center—is proportional to the centrality or commonality of connection with the node in question. **H** shows how the angle between any two nodes $\theta$ is proportional to the distance between the nodes. **I** illustrates this polar coordinate system for several plotted points.

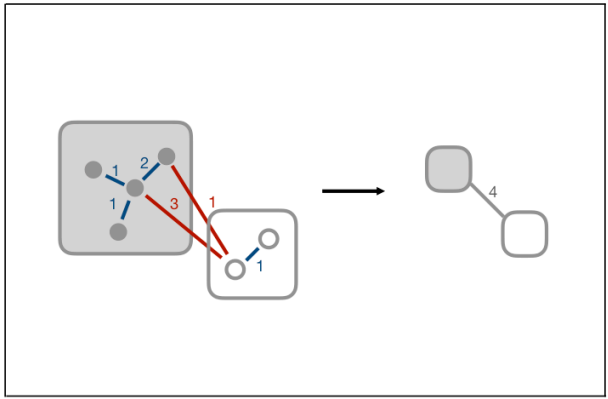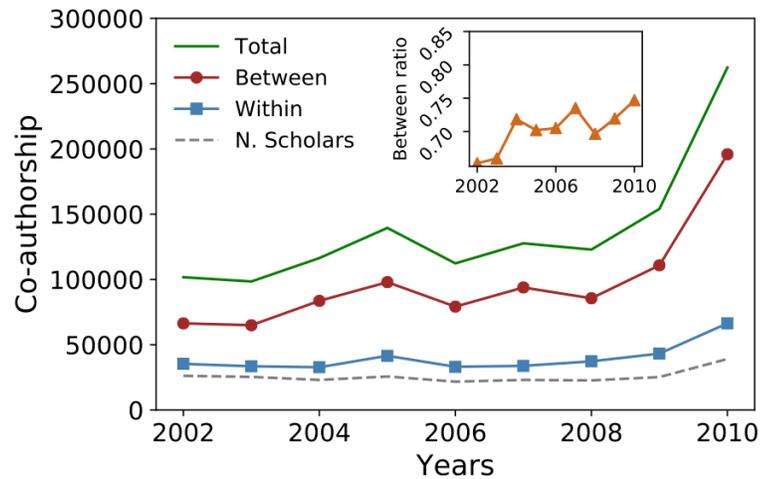

**Figure 2. The construction of institutional networks.** Panel A shows the co-authorship between five scholars (dots) from two universities (rectangles). Co-authorship within universities is colored in blue and co-authorship between universities is colored in red. Weights represent the number of co-authorships in a given year. When we construct institutional collaborations we keep only the co-authorship between universities. Panel **B** shows the yearly counts of the number of scholars (the gray, dotted line), the total co-authorship between these scholars (green), which is separated into co-authorship between (red) or within (blue) institutions. Panel **B** reveals that the number of between coauthorships is proportional to the number of total coauthorships and so highly descriptive of the entire pattern of coauthoring ties.

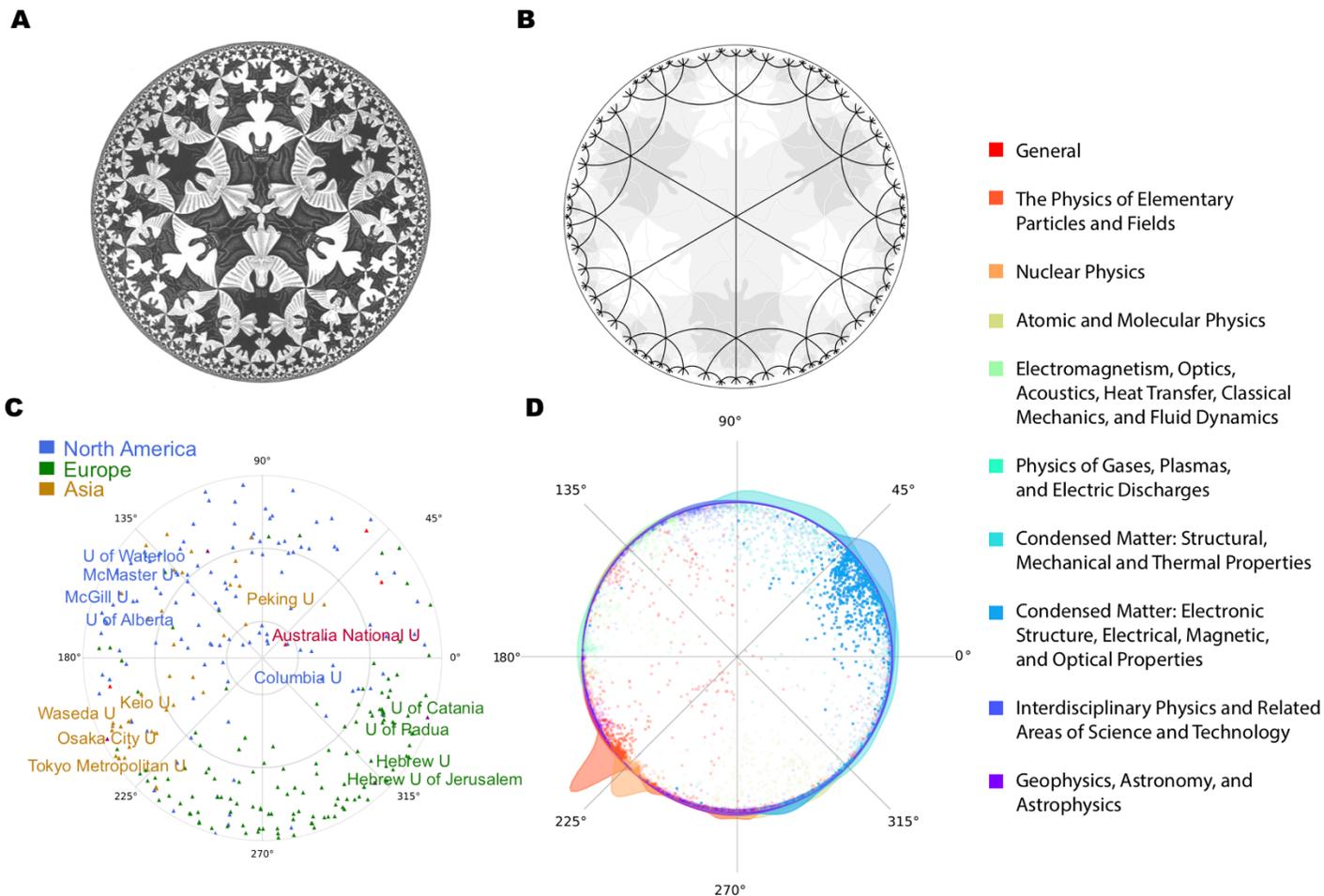

**Figure 3.** Panel A and B present M.C. Escher's print *Circle Limit IV* and the underlying the Poincaré disk model, recited and recreated from (Dunham and Others 2009). The space becomes exponentially denser as the radius increases. Panel C shows the hyperbolic embedding of the institution collaboration network in 2011 using the the Poincaré disk model. The displayed 290 institutions (triangles) are colored by regions: blue for North America, green for Europe, orange for Asia, red for Australia, and purple for South America. Panel D shows the hyperbolic embedding of the aggregated PACS code co-occurrence network using the the Poincaré disk model. The analyzed 5,819 PACS codes (dots) are colored by the 10 subfields to which they belong. Gaussian kernel density estimation is used to visualize the concentration of the angles of PACS codes in each subfield (see Figure A5 for details). To compare across subfields, we also rescale the estimated distributions such that the area covered by the distribution curve is not unity, but proportional to the total number of papers published by 290 institutions in this subfield from 2002-2011.

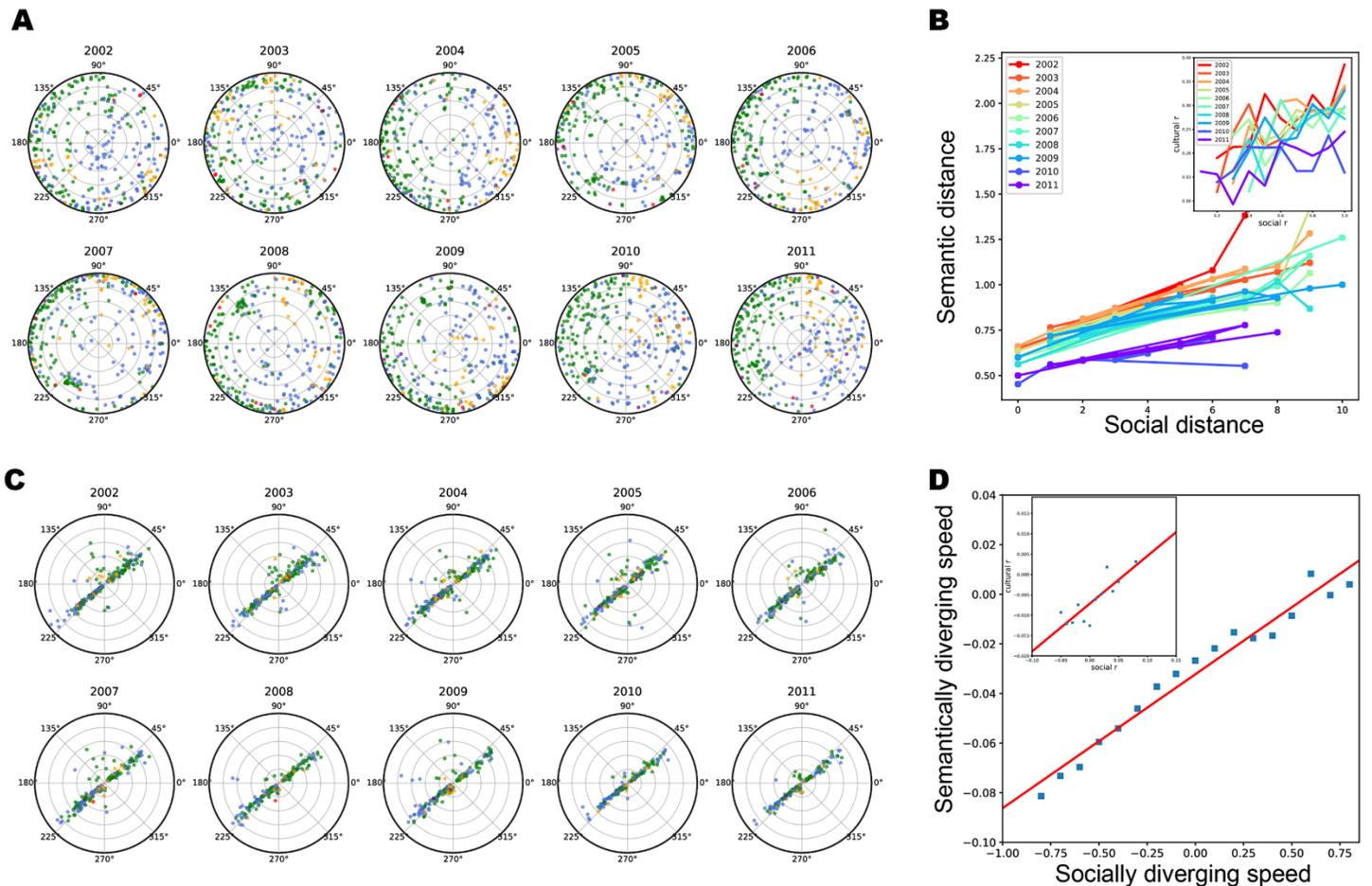

**Figure 4.** (A) The hyperbolic embedding of annual university physics collaboration networks (2002-2011). We embed each annual collaboration network into a hyperbolic space and project them on Poincaré disks, such that each institution $i$ has two parameters in the hyperbolic social space, angles $\theta_{si}$ and radius $r_{si}$. The universities (dots) are colored by region, blue for North America, green for Europe, orange for Asia, red for Australia, and purple for South America. (C) Annual projection of institutions in the hyperbolic space of physics topics. We embed the PACS code co-occurrence network into Poincaré disks as shown in Figure 3D, and then calculate the position of each university in a given year from the position of their PACS codes, such that each institution $i$ has two parameters in the culture space, angles and radius $r_{ci}$. The manner by which we aggregate the PACS code parameters is illustrated in Figure A5. (B) The correlation between social and semantic distances. We calculate the hyperbolic distance between every pair of universities in both Social and Semantic space as $C_{ij}$ and $S_{ij}$ for all pairs of institutions in each year, then plot $C_{ij}$ against $S_{ij}$ across years (data is binned). The inset shows the correlation between social and semantic hyperbolic distances at the university level, i.e., we calculate the average hyperbolic social and semantic distances from one institution to all the other institutions $1/n \sum_{j=1}^{n} S_{ij}$ and $1/n \sum_{j=1}^{n} C_{ij}$, and plot them against on another. (D) The correlation between the social and semantic time series, social distance $S_{ij}(t)$ and semantic distance $C_{ij}(t)$. For each pair of institutions, we use an OLS regression to fit the slopes $B_{sij}$ (social diverging speed) of $S_{ij}(t)$ against year $t$ and $B_{cij}$ (semantic diverging speed) of $S_{ij}(t)$ against years $t$. The large, positive slopes imply that two institutions are moving toward and away from each other in the corresponding spaces. We find that $B_{sij}$ and $B_{cij}$ are positively correlated, with a Pearson correlation coefficient equal to 0.98 (P-value < 0.001). Similarly, in the inset we show that the slopes of the time series $r_{si}(t)$ and $r_{ci}(t)$ are also correlated, with a Pearson correlation coefficient equals 0.85 (P-value < 0.001).

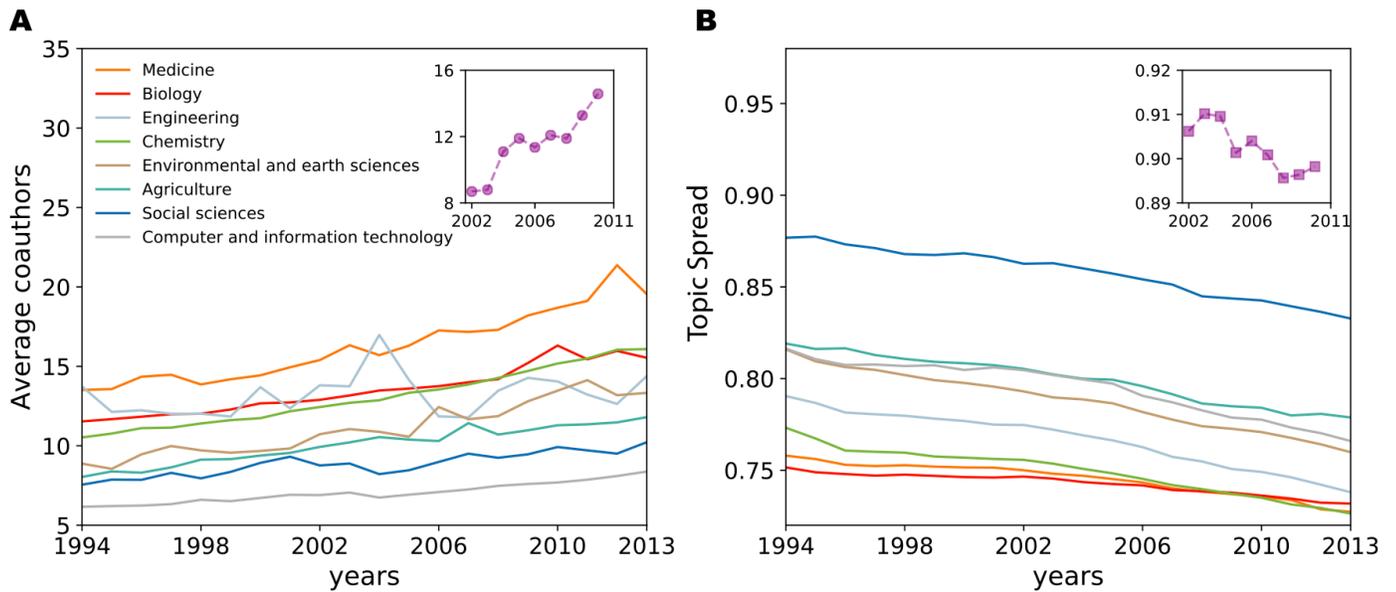

**Figure 5. The densification of collaboration networks and the convergence of topics and wording.** We selected and analyzed 13,178,911 journal articles from our WoS dataset across seven fields during 1994-2013. **A.** The average number of coauthors increases over time for a majority of fields. The fields are based on WoS meta data of journal classifications. To count the average number of coauthors, we sum over the number of papers, weighted by number of coauthors in the paper, for each name-disambiguated scholar and average over all scholars. The inset shows the increase of coauthors for APS data. **B.** The spread (entropy) of words from article abstracts decreases over time for a majority of fields. We select 3,510,489 words that occur twice or more in the abstract to represent a research article (commonly used words are removed). To control the word vocabulary length difference across years we normalize the calculated entropy by its theoretical maximum Log(N), in which N is the vocabulary length. The inset shows the decrease in entropy of the PACS codes.

# Appendix

As we will show below, a Poincaré disk can accurately represent a branching tree structure, and higher dimensional hyperbolic spaces can capture hierarchies with multiple heritage, such as directed acyclic graphs. As a result, negative curvature and diverging, branching structure of hyperbolic space enables us to model the hierarchy and sparse, bridging structures common to many realistic social and semantic networks in far fewer dimensions than a comparable number of Euclidean dimensions

**Physics, Hyperbolic Geometry and Structural Representation**

It is notable that hyperbolic geometry was connected with physics from inception. Nikolai Lobachevsky, one of hyperbolic geometry's 19th Century discoverers, argued that the universe might not be flat, but rather hyperbolic (Bonola 1955), and Einstein's general relativity intensified interest in the curvature of the universe by positing that mass and energy bend spacetime. In general relativity, mass and energy can be used to determine the universe's curvature ($\Omega$) as the average density of the universe divided by the mass energy required for it to be flat. Despite recent estimates that put $\Omega$ at approximately 0 (Biron 2015), spacetime in the vicinity of massive bodies such as stars, solar systems and galaxies is positively curved, and space between massive bodies, in dark energy dominated areas[11], is negatively curved (Krioukov et al. 2012).

Statistical physicists, computer and network scientists first applied hyperbolic geometry as a compact representation of social networks (Papadopoulos et al. 2012; Krioukov et al. 2010) and semantic hierarchies (Nickel and Kiela 2017; Chamberlain, Clough, and Deisenroth 2017). A Poincaré disk can represent a branching tree structure without distortion, and higher dimensional hyperbolic spaces can perfectly capture hierarchies with multiple heritage, such as directed acyclic graphs. As a result, negative curvature and diverging, branching structure of hyperbolic space enables us to model the hierarchy and sparse, bridging structures common to many realistic social and semantic networks (Nickel and Kiela 2017).

**Euclidean geometry v.s. Hyperbolic geometry**

---

[11] As dark energy is constant over the universe, regions with little mass are proportionally "dominated" by dark energy.

*Euclidean geometry* was constructed to obey Euclid's fifth postulate, which states that within a two-dimensional plane, for any given line ℓ and point p not on ℓ, there exists exactly one line through p that does not intersect ℓ. The result is a space dimensionalized by "straight" vectors and "flat" planes having zero curvature, such that a drawn triangle's angles add up to 180° and the Pythagorean theorem holds.[12] Curved geometries including both elliptical and hyperbolic geometries violate Euclid's fifth. A 2-dimensional elliptical geometry, for example, has positive curvature and can describe the surface of the earth, such that all lines through p intersect ℓ, and any two lines perpendicular to ℓ intersect at a "pole". The sum of a triangle's interior angles on an elliptical surface is always greater than 180°.[13]

Hyperbolic space, by contrast, has negative curvature, such that infinitely many lines may go through *p* without intersecting ℓ. A saddle, mountain pass, coral reef, or crocheted frill furnish examples of negatively curved surfaces. A triangle etched upon them will have angles that sum to less than 180°. As two intersecting lines diverge at a constant rate in Euclidean geometry (e.g., the Pythagorean theorem), and converge in elliptical geometry, they diverge exponentially in hyperbolic geometry such that there is "more space" in a 2-dimensional hyperbolic disc than a Euclidean circle or elliptical globe. In Euclidean geometry, the circumference of a circle with radius *r* equals *2πr*, but in hyperbolic geometry it is always more and elliptical geometry always less. Figure 3, panel A presents M.C. Escher's print *Circle Limit IV* and panel B shares the underlying Poincaré disk model, the projection of a 2-dimensional hyperbolic space onto a 2-dimensional Euclidean space, cited and recreated from (Dunham and Others 2009). Escher's print alternatively tessellates angels and devils on the disk, *ad infinitum*, and reveals how the space becomes exponentially more dense as the radius increases.

The inner products in Poincaré disk and Euclidean space are defined as follows. For two points $x = (r_x, \theta_x)$, $y = (r_y, \theta_y)$ in the Poincaré disk using polar coordinates, the hyperbolic inner product of $x, y$ is $\langle x,y \rangle = \|x\| \|y\| \cos(\theta_x - \theta_y) = 4 \text{ arctanh } r_x \text{ arctanh } r_y \cos(\theta_x - \theta_y)$. In a Euclidean space, the inner product of $x, y$ is defined as $\langle x,y \rangle = \|x\| \|y\| \cos(\theta_x - \theta_y) = 4 r_x r_y \cos(\theta_x - \theta_y)$.

---

[12] The Pythagorean theorem states that the square of the hypotenuse of a right triangle (the side opposite the right angle) is equal to the sum of the squares of the other two sides.

[13] Consider a triangle etched onto earth's surface with one side stretching along a quarter of the equator and the other two sides meeting at the South Pole. All three angles will equal approximately 90°, summing to 270°.

**The Information Theory of Cultural Collapse**

Shannon's information theory has recently been used in cultural models of meaning (Martin and Lee 2018) and furnishes a formal model of communication and information (Shannon and Weaver 1963) that illuminates the theoretical and observed relationship between social connection and cultural contraction. Communication plays a central role in information theory and can be equated with social connections or pre-existing "information channels" required to exchange messages. Information, in Shannon's theory, is modeled as message passed between sender $X$ and receiver $Y$ that resolve the receiver's uncertainty about the sender's communicated meaning. Messages may express and transmit cultural forms including tastes, expectations, ideas, behaviors and artifacts (Carley 1986, 1991; DiMaggio 1987; Erickson 1988; Kandel 1978; Krackhardt and Kilduff 1990; Krohn 1986). Information $I$ is equivalent to the surprise of experiencing a given message or cultural communication, such that an improbable or unexpected message has more information than an anticipated one and is experienced by the receiver with greater surprise. Information, then, is the news quality of a message. If senders and receivers of information mingle through repeated back-and-forth communication, however, the *mutual information $I(X;Y)$* of their messages—the information common to both $X$ and $Y$, modeled as two random variables—increases. This increase in mutual information necessitates a drop in *joint entropy $H(X,Y)$*—the average information produced by both actors communicating together (see Figure A1, top panel). This relationship assumes that individuals exhibit a conserved number of cultural forms and that what forms they possess, they express (Leydesdorff and Ivanova 2014).

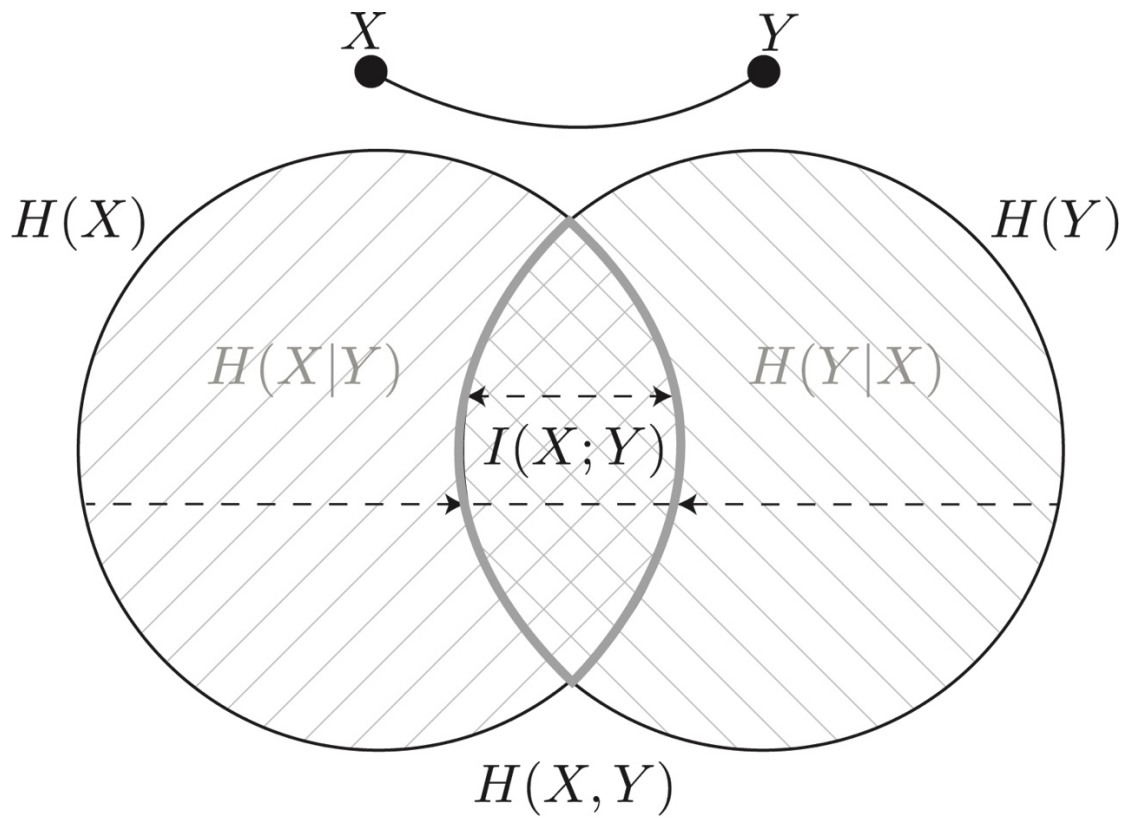
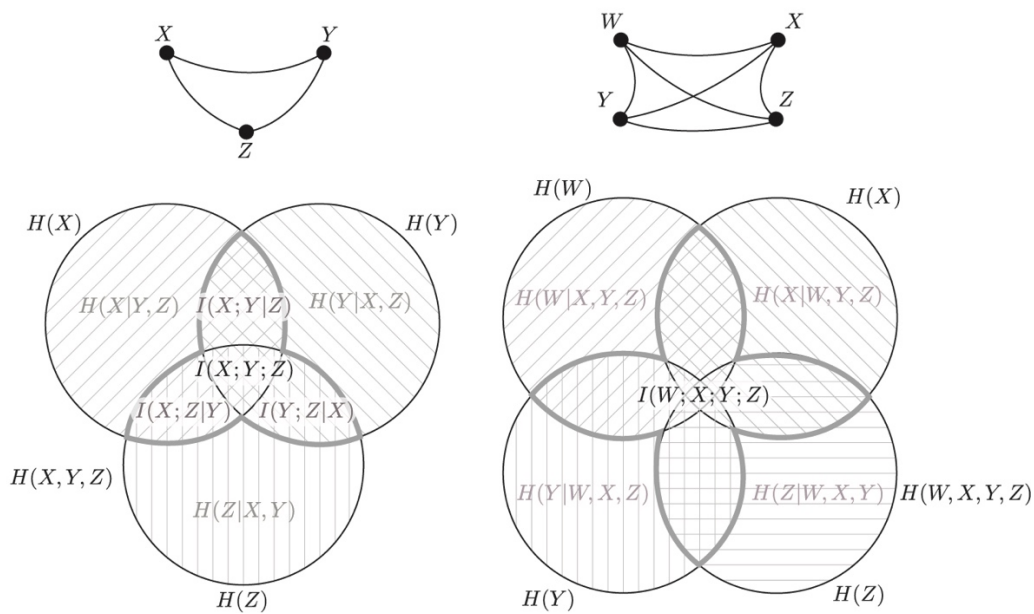

**Figure A1. Venn diagrams for information measures calculated for culturally correlated actors W, X, Y and Z.** For the simple {X, Y} network, H(X) constitutes the entropy or average information associated with X's cultural forms and H(Y) the entropy of Y's cultural forms. As their relationship and similarity increases, their mutual information I(X;Y) rises and their joint entropy H(X,Y)—the total information in the system—shrinks. H(X|Y) is the conditional entropy of X given Y, or the average cultural information or surprise another actor A would experience when communicating with X after having communicated with Y. The graphs and Venn diagrams at the bottom of the figure illustrate the information scenario for clustered networks {X, Y, Z} and {W, X, Y, Z} (with a missing W, Z intersection for the {W,X,Y,Z} network), where the joint entropy shrinks linearly with the union of all multivariate mutual informations (see footnote 3).

The pattern for two communicating actors generalizes to an arbitrary number of communicators linked via network. In larger networks, the inverse relationship holds between the multivariate mutual information and multivariate joint entropy, with some slight technical amendment. In the case of many variables—or individuals in society—the multivariate mutual information can be positive or negative, and so for the identity to remain precisely equivalent in a larger network and that of a single dyad, the joint entropy shrinks linearly with growth of the union between (1) the multivariate mutual information (e.g., $I(X;Y;Z)$ in a three-person network; see lower left panel in Figure A1), and (2) all conditional mutual information measures (e.g., $I(X;Y|Z)$, $I(X;Z|Y)$, $I(Y;Z|X)$ in a three-person network; see lower left panel in Figure A1). As a result, when increased channels of communication are unleashed by a denser pattern of social relationships with an arbitrary number of individuals, the cultural information passed between them shrinks in direct proportion to their shared or mutual information.

This information theoretic rendering of the relationship between structure and culture reinforces the mutually constitutive relationship between individual-level innovation unleashed by novel social connections and communication (Burt 2004) and the system-level diversity depressed by it (March 1991). Connecting disconnected societies enables cultural mixing in the short-term, but through competition and selective cultural extinction, future periods have fewer cultural forms to mix. The more socially and culturally distant the groups connected, the greater the immediate innovation potential for the socio-cultural "arbitrageurs" that connect them, but the larger decrease in system-level diversity that results from putting more cultural forms in contact and competition. As mentioned in the text, this suggests that the second law of thermodynamics may have a manifestation in culture, such that the distribution of cultural forms in a social system becomes more *even* as social relationships increase the mutual exchange of information.

**Name disambiguation of authors and institutions**

In order to disambiguate WoS authors we use a hybrid algorithm. For each name (including family name and initials), we construct a network of relevant papers connected on the basis of a similarity measure that considers shared co-authors, references and citations. Disconnected components of this network are assumed to correspond to distinct authors (Wu, Wang, and Evans 2017). We obtain 10 million scholars who contributed to 22 million papers. Eighty-five per cent of these scholars contributed to three or more papers, and 44% to four or more. We use the 2017 Open Researcher and Contributor ID (ORCID) dataset to validate the name disambiguation results, and find that precision is 78% and recall is 86% among the 118,094 ORCID scholars with three or more papers. We also test the results using a dataset of 31,070 Chinese scholars and 253,786 papers retrieved from the project outcome reports of research funded by National Natural Science Foundation of China with precision 79%, and recall 65%.[14]

APS publication data is a freely accessible online (https://journals.aps.org/datasets), containing all meta-data for articles published in *Physical Review Letters*, *Physical Review* (*A, B, C, D, E and X*), and *Review of Modern Physics*, some of which predate the APS, going back to 1893. In order to analyze collaborations between scholars and institutions, we first needed to disambiguate author and institution names: the same name may refer to multiple people, and distinct affiliation names reference the same institution. To extract unique author names, we matched the APS meta-data with WoS data using the DOIs (Digital Object Identifier) of papers published during 2001-2013. From this disambiguated data, we successfully identified 372,495 authors from the analyzed 359,395 APS papers.

To resolve ambiguity among institution names, we extracted affiliation names and fed them into the Google Geolocation API to identify whether the same latitudes and longitudes are returned. Google Geolocation provides results within 25 feet if the address translation is correct. Leveraging this approach, we merged 330,560 institution names into 55,317 unique ones.

**Modeling complex social and semantic networks using the Poincaré disk model**

The initial algorithm used to represent social networks in hyperbolic space was slow and inefficient (Papadopoulos, Psomas, and Krioukov 2015). The primary bottleneck involved the Hamiltonian Monte Carlo method used to estimate node positions, which repeatedly calculated the gradient or fit between network and geometric embedding and became burdensome as datasets scaled up. This changed dramatically when computer scientists discovered that artificial neural networks were powerful tools for embedding optimization. Moreover, analysts have demonstrated that low dimensional hyperbolic embedding improve upon higher dimensional Euclidean embedding for many tasks, including predicting the collaboration between scientists in weighted undirected collaboration networks (Nickel and Kiela 2017; Chamberlain, Clough, and Deisenroth 2017).

---

[14] "Precision" here is defined as the percentage of cases that are "correctly identified"(true positive), and "Recall" is the percentage of total relevant cases that are actually achieved by the method.

In Figure A2, we graph the institution collaboration network over time, with all 290 institutions (dots) colored by region: blue for North America, green for Europe, orange for Asia, red for Australia, and purple for South America. Edges represent the co-authoring relationship between scholars from each institution. Between 2002 and 2011, the number of edges increased from 3,331 to 7,647, and the total weight increased from 29,117 to 94,336 over the same period.

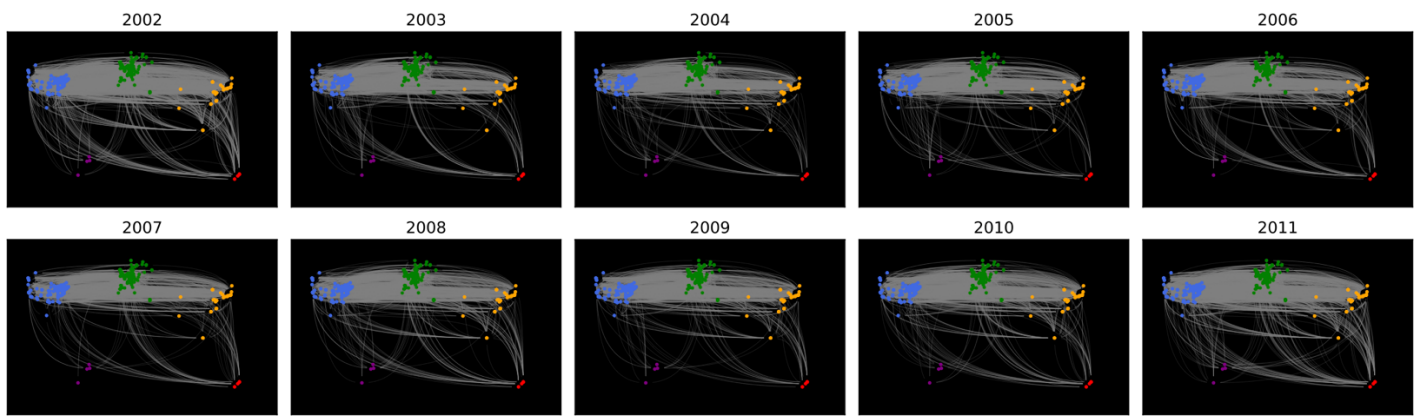

**Figure A2. Institution collaboration network, 2002-2011.**

In Figure A3 we illustrate the movement of institutions in collaboration space over time. We pick Columbia University as the focal institution (the red triangle) and investigate the temporal evolution of its distance to two collaborating institutions, the University of Notre Dame (the orange rectangle) and University of Southern California (the blue rectangle). Over time, the University of Notre Dame moves towards Columbia University, whereas the University of Southern California moves away from it.

We also highlight the top ten institutions closest to in Columbia University in 2002 (purple dots) and their position in the following years to provide visual guidance for distance in the hyperbolic space, which differ substantially from 2D Euclidean distance.

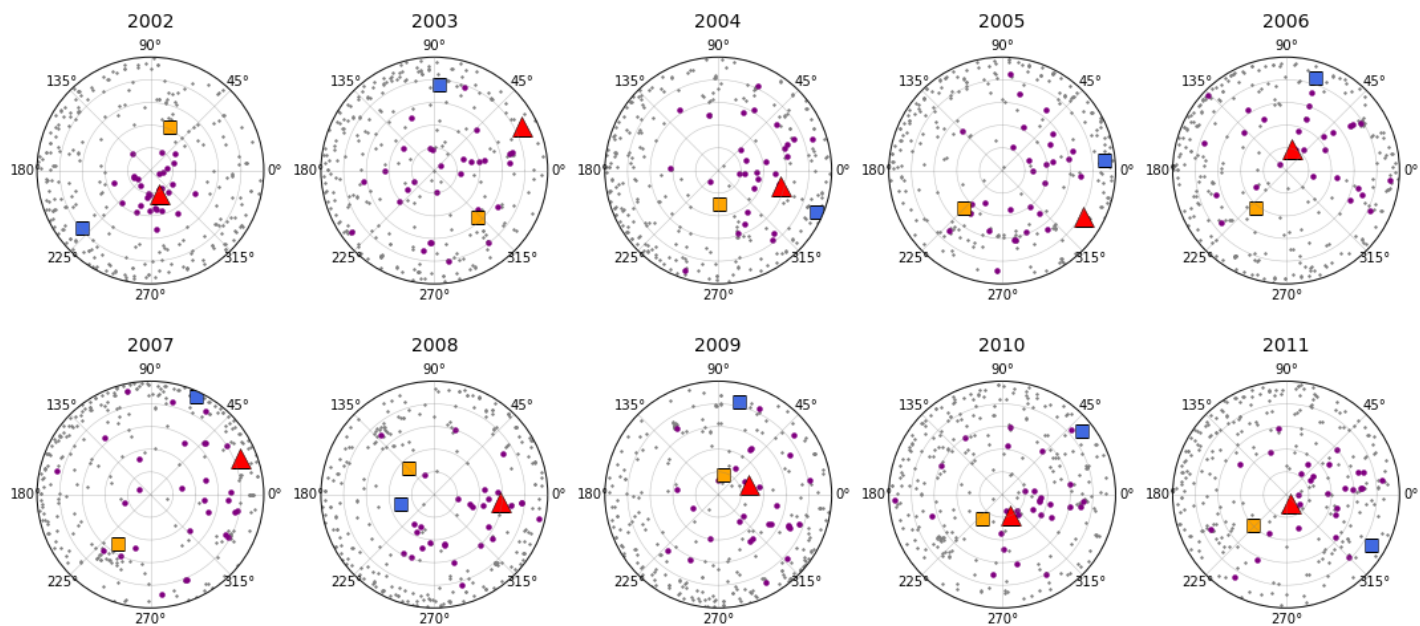

**Figure A3. Movement of institutions in collaboration space, 2002-2011.**

In Figure A4, we illustrate the movement of institutions in the semantic space of physics over time. We constructed the culture space as shown in Figure 3D, and display all 5,819 PACS codes in the panels above as gray dots. Next, we calculated the representative locations of Columbia (red triangle), Notre Dame (orange rectangle) and University of South California (blue rectangle), as shown in Figure A3, but within the culture space based on the positions of their PACS codes using the method introduced below and illustrated in Figure A5. We also fitted Gaussian kernel density curves to show the concentrate and spread of PACS codes for each of these three institutions.

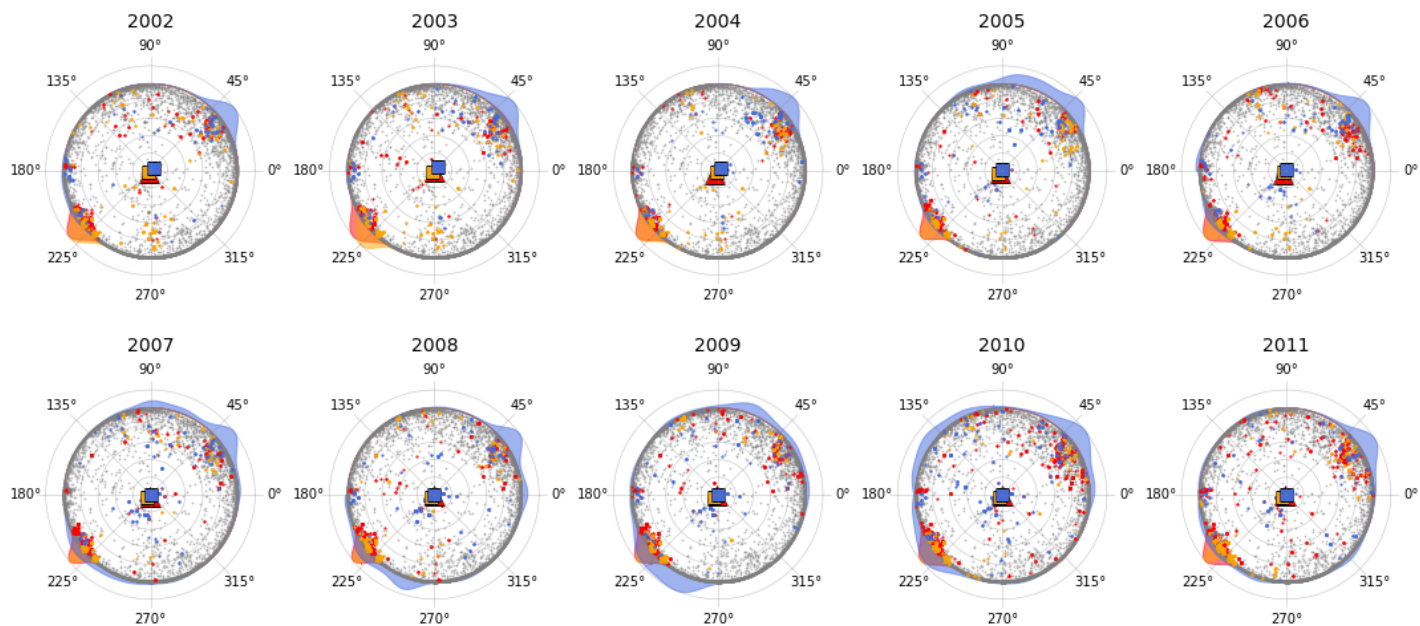

**Figure A4. Movement of institutions in the "semantic space" of physics topics, 2002-2011.**

In Figure A5, we illustrate the empirical distribution of angles and radius for PACS codes from Figure 3D. Each row shows the 5,819 PACS codes from one of ten subfields, which are colored following the same scheme as Figure 3D. Due to the size limitations of the figure, abbreviated names of the subfields are used. The first column shows the two estimated Poincaré disk parameters, angles $\theta_i$ (the x-axis) and radius $1-r_i$ (the y-axis; for figure readability rather than $r_i$) of the $i$th PACS code. The second column shows the empirical distribution of angles $\theta_i$ and corresponding Gaussian kernel density estimation. The red lines show the peak of the angle distribution we select to represent the summary research direction of subfields or institutions based on their PACS codes. The third column shows the empirical distributions of radius $1-r_i$. We found that due to the asymmetric property of the hyperbolic space, the values of most $r_i$ approach 1. As such, neither mean nor median can characterize the difference in hierarchy between two universities or subfields. Therefore we take the extreme values by calculating the average of the ten smallest $r_i$ to represent the university or subfield hierarchy.

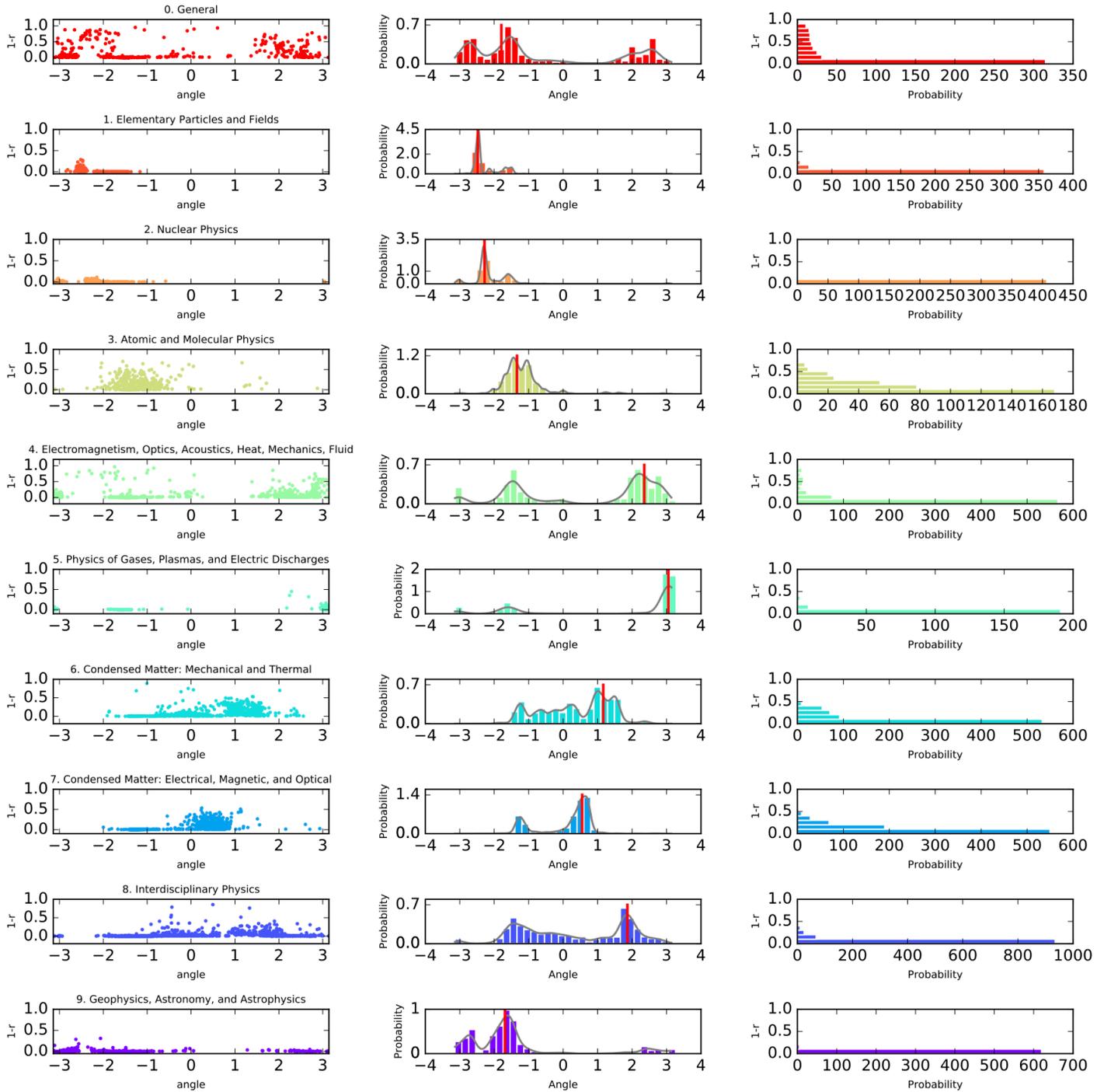

**Figure A5. Illustration of the empirical distribution of angles and radius for all PACS codes from Figure 3D.** we displayed the probability *density* function and not probability *mass* function. The former does not sum up to one, but these two functions are of the same shape.

After we constructed a university collaboration network, we also measured the density of social collaborative connections. Following convention, we define social density as the ratio of observed network edges to the maximum number possible (Wasserman and Faust 1994). We denote the set of weighted edges connecting the $i$th node into the global institutional collaboration network as $E_i$, which can be further separated into $E_i = E_i^{in} + E_i^{out}$ where $E_i^{in}$ denotes self loops and $E_i^{out}$ edges connecting node *i* with other nodes (Palla, Barabási, and Vicsek 2007). For institution *i*, we defined a weighted clustering coefficient:

$$c_i = \sum_{kh} (|E_k^{out} \cap E_h^{out}| \, |E_i^{out} \cap E_k^{out}| \, |E_i^{out} \cap E_h^{out}|^{1/3} / (max(w) * |E_i^{out}| * (|E_i^{out}|-1)))$$

where *k* and *h* include all unique pairs of institutions that constitute closed triplets with *i* and *max(w)* represents the maximum weight of *i*'s ego network. We use $c_i$ to measure social density as an alternative to measures defined in the hyperbolic social space. We also calculated the entropy of PACS codes as a measure of semantic diversity. In Figure A6 we find that social density and semantic hierarchy are positively correlated and that social density and semantic diversity are negatively correlated. Therefore, our expectation that the broad association between social interaction and semantic contraction appears to hold for 21st Century Physics.

If position in semantic hierarchy is defined as radius $r_{ci}$ and semantic diversity $S_{ci}$ as the entropy over their PACS code angles on the Poincaré disk, then the two main figures show the temporal evolution of $c_i$, $r_{ci}$, and $S_{ci}$ for two extreme cases. The left panel shows how social density ($c_i$, the left y-axis) and semantic hierarchy ($r_{ci}$, the right y-axis) move in the same direction. More specifically, both $c_i$ (dotted red line) and $r_{ci}$ (solid red line) for the University of Innsbruck decreases over time. We plot these as ranks and not original values and also adjust the scale of ranks to provide visual guidance for comparing trends. By contrast, both $c_i$ (dotted blue line) and $r_{ci}$ (solid blue line) for the Naval Research Lab increase over time. The right panel shows how social density ($c_i$, the left y-axis) and semantic diversity ($S_{ci}$, the right y-axis) go in the opposite direction. The $c_i$ for the University of Innsbruck decreases over time (dotted red line), whereas its PACS code entropy increases (solid red line). By contrast, the $c_i$ of Johns Hopkins University increases over time (dotted blue line), but its PACS code entropy decreases (solid blue line). The two insets confirm that the full sample social density and semantic hierarchy are positively correlated and the full sample social density and semantic diversity are negatively correlated.

In Figure A10, we explore the directional prediction of social and semantic distance, and their component parts—radius and angular distance—over time, by lagging each variable by one year and predicting the others. Even though it takes time for collaboration to yield a research article with particular PACS codes, both collaboration and PACS codes are assessed from publications, and so predicting future collaborations from present PACS codes indicates the likelihood that a similar publishing focus at time *t* will attract like-minded physicists to work together at time *t* +1. Conversely, predicting future PACS codes from current collaborations indicates the likelihood that collaborations with a partner institution at time *t* anticipates a convergence in publishing focus at time *t* +1. When we perform these regressions between each pair of universities, we discover the degree to which social centralization leads to convergence, and its converse. When we perform them for each university and all other universities, we unearth the extent to which social centralization leads to semantic contraction. Figure A10 summarize all of these Granger causal regressions. The number decorating each arrow in the graph represents the percentage of significant Granger causal regression estimates that post a positive influence.

From these figures, we see that the apparent association causality between social and semantic distances is bidirectional, suggesting that both socialization and selection are driving the attention of universities' physicists. Nevertheless, present collaborative distance drives future semantic or topical distance slightly more consistently and significantly than its converse. 67% of all regressions predicting next year's topic distance from each pair of universities' current social distance post a positive effect, compared with only 57% that do so in the opposite direction. By contrast, 89% of all regressions predicting next year's topic distance from current global collaborations between a university and all others post a positive effect, compared with 79% from collaboration to shared focus. These patterns provide moderate evidence that social centralization leads to semantic convergence between pairs of universities, but strong evidence that more global social centralizations lead to semantic contraction.

When we decompose the hyperbolic distance between pairs of universities, and between each university and all others, into the difference between their angle ("divergence") and radius ("hierarchy"), we find that hierarchical distance in the social or semantic domain is always a more consistent predictor of hyperbolic distance in the other domain than angular distance. This suggests a more consistent sensitivity of social and semantic distances to differences in the centrality of a university's physicists in the alternative space. Greater centrality provides universities and their physicists with a wider menu of options, and less centrality gives them far fewer choices regarding what they can do or where they can go next. As a result, differences in social centrality more consistently magnify differences in semantic distance, and the converse, than differences in angle, which represents a university's specific topical focus or collaboration cluster. This provides substantial support for the importance of using a hyperbolic representation that highlights hierarchy for the social and semantic systems characterized by it.

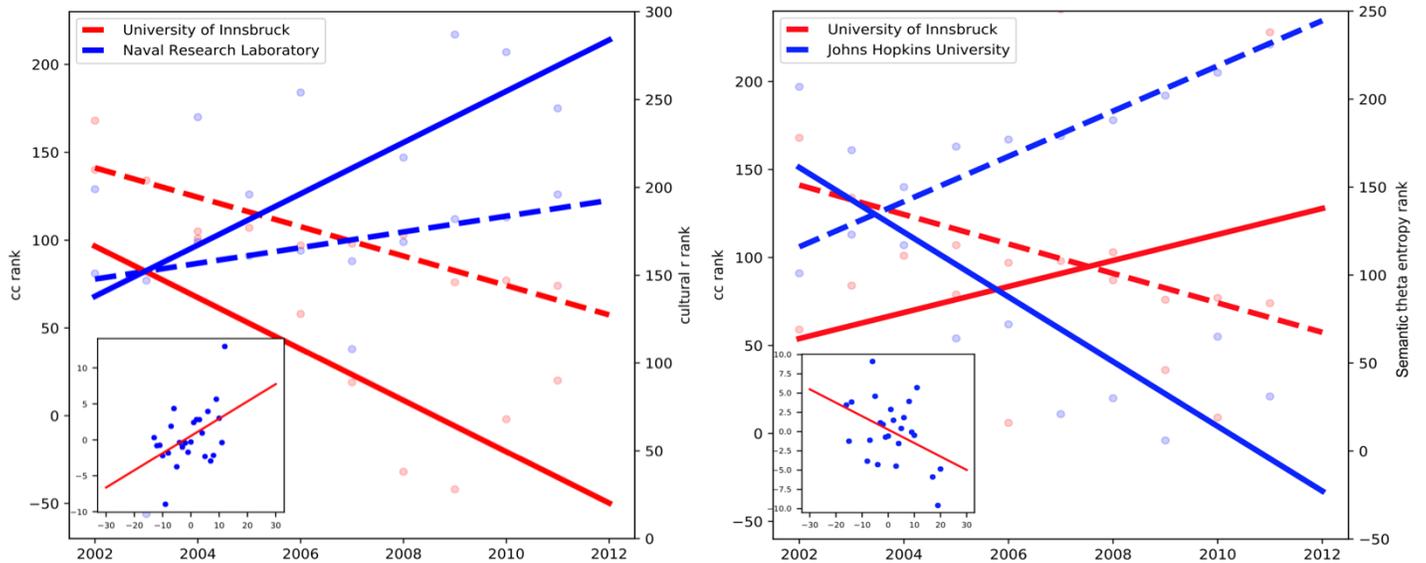

**Figure A6. Correlation between social density, semantic hierarchy, and semantic diversity.**

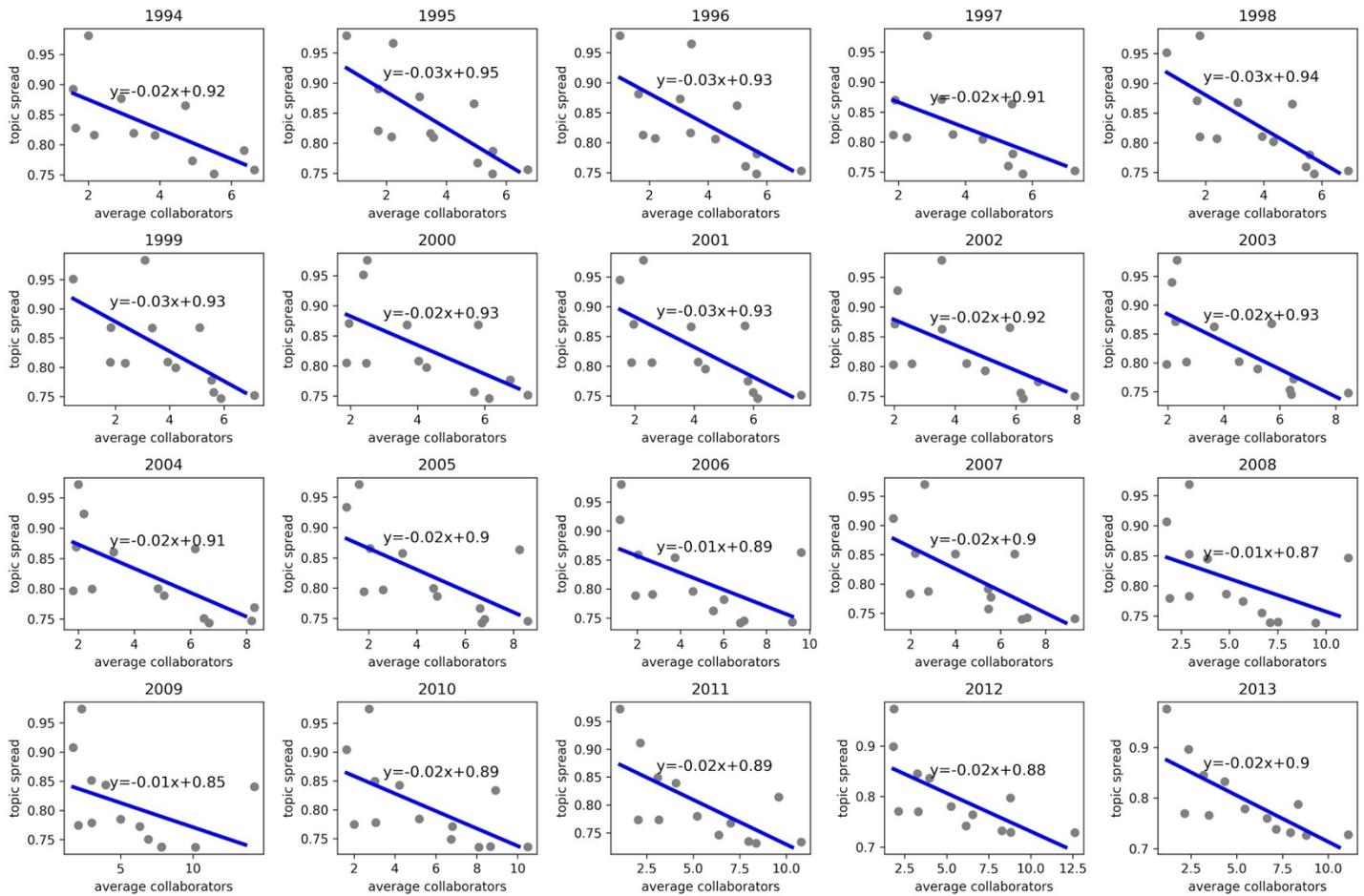

**Figure A7. Correlation between social densification and topic convergence across 13 fields, 1994-2013.** In each subplot, a data point represents a field in science. The average collaborators (x-axis) and the topic spread (y-axis) are computed in the same way as in Figure 5. The OLS regressions are used to fit the data points, a majority (85%) of the regressions are significant (p < 0.05) and all the regressions are marginally significant (p < 0.1). The regression coefficients (the slopes of blue lines) vary from -0.01 to -0.03.

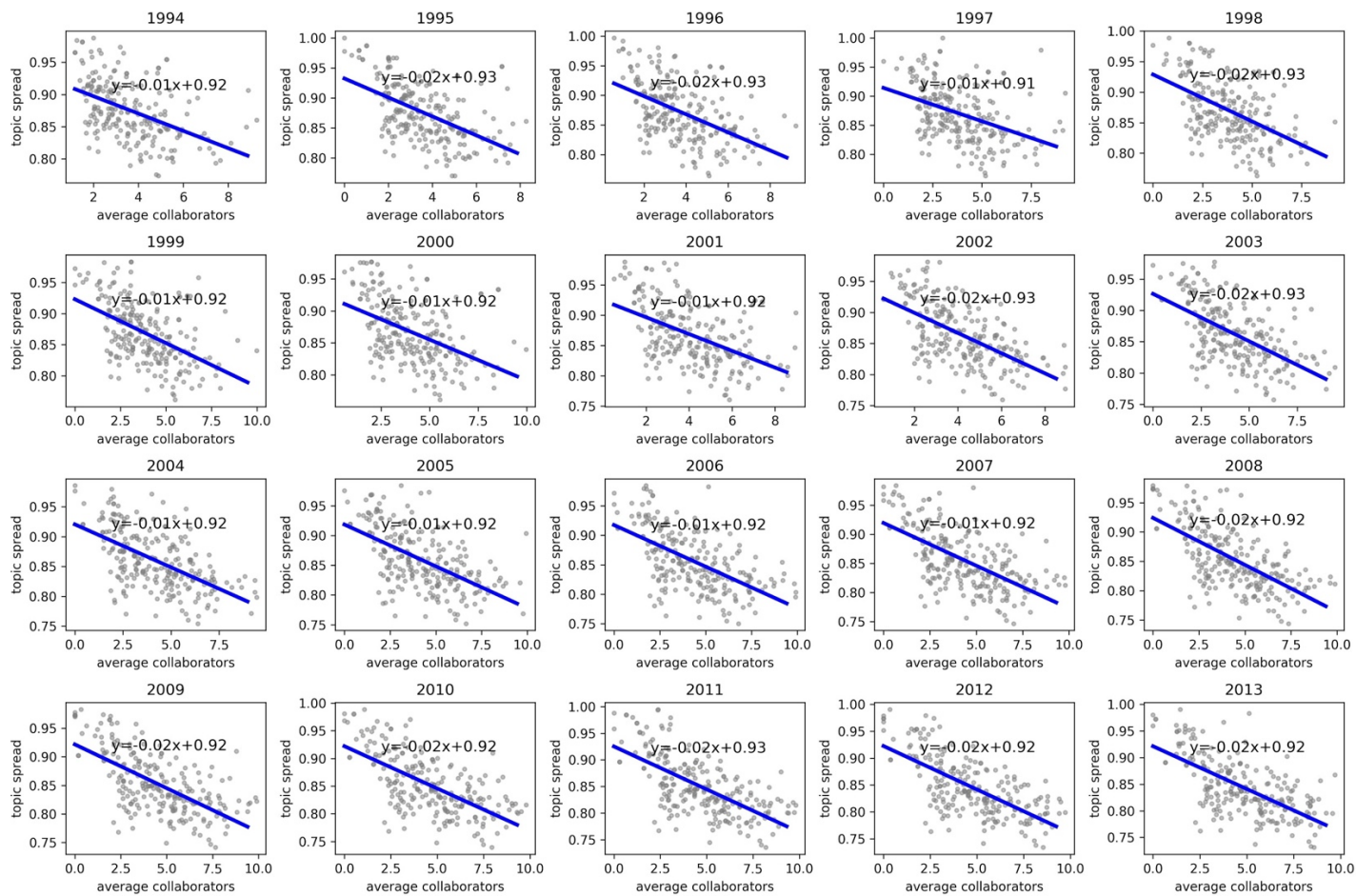

**Figure A8. Correlation between social densification and topic convergence across more than 200 subfields, 1994-2013.** In each subplot, a data point represents a subfield in science. The subfields are based on WoS meta data of journal classifications. The average collaborators (x-axis) and the topic spread (y-axis) are computed in the same way as in Figure 5. The OLS regressions are used to fit the data points, all the regressions are significant ($p < 0.05$). The regression coefficients (the slopes of blue lines) vary from -0.01 to -0.02.

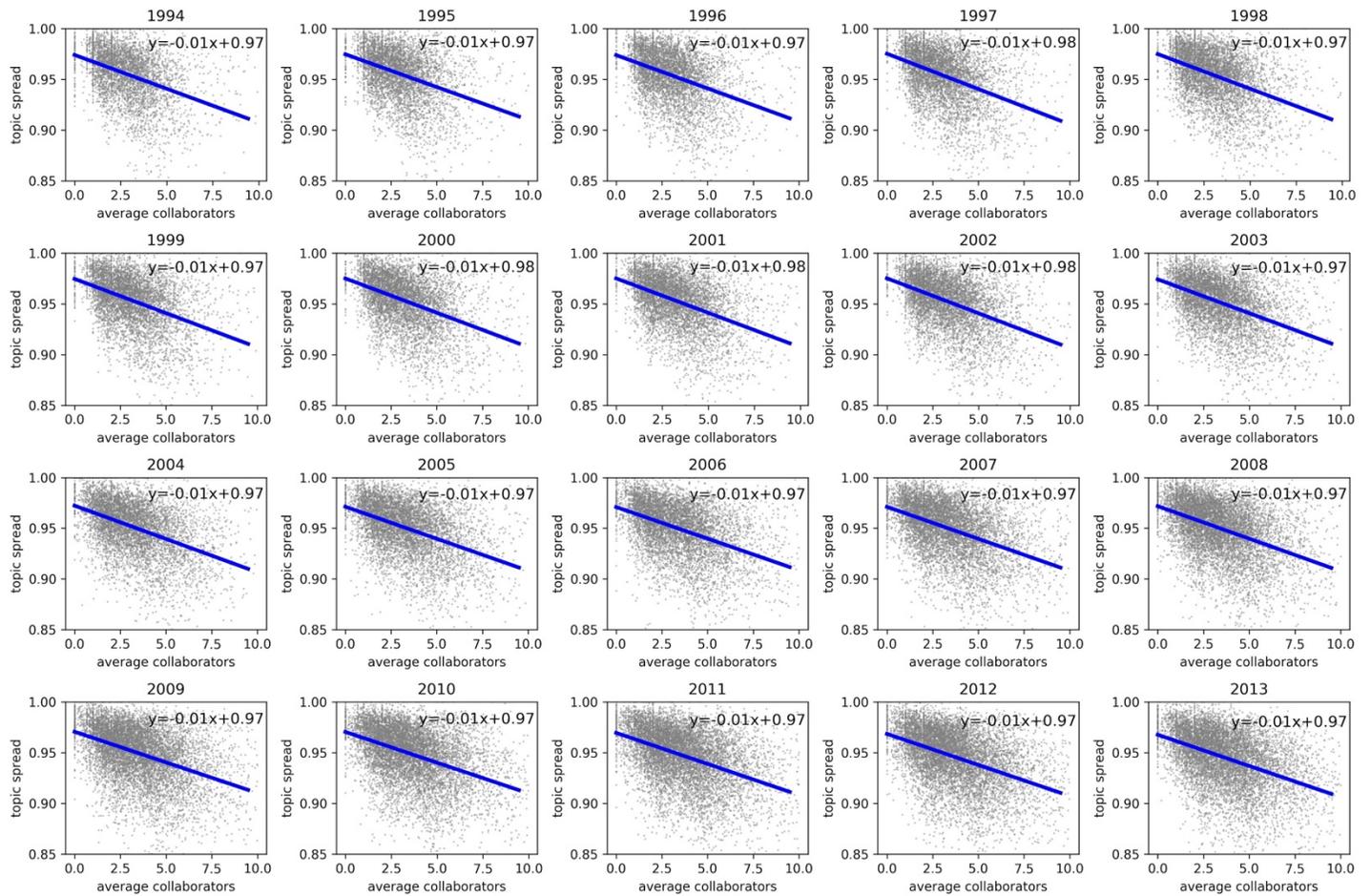

**Figure A9. Correlation between social densification and topic convergence across over 10,000 journals, 1994-2013.** In each subplot, a data point represents a journal. The average collaborators (x-axis) and the topic spread (y-axis) are computed in the same way as in Figure 5. The OLS regressions are used to fit the data points, all the regressions are significant ($p < 0.05$). The regression coefficients (the slopes of blue lines) for all subplots are -0.01.

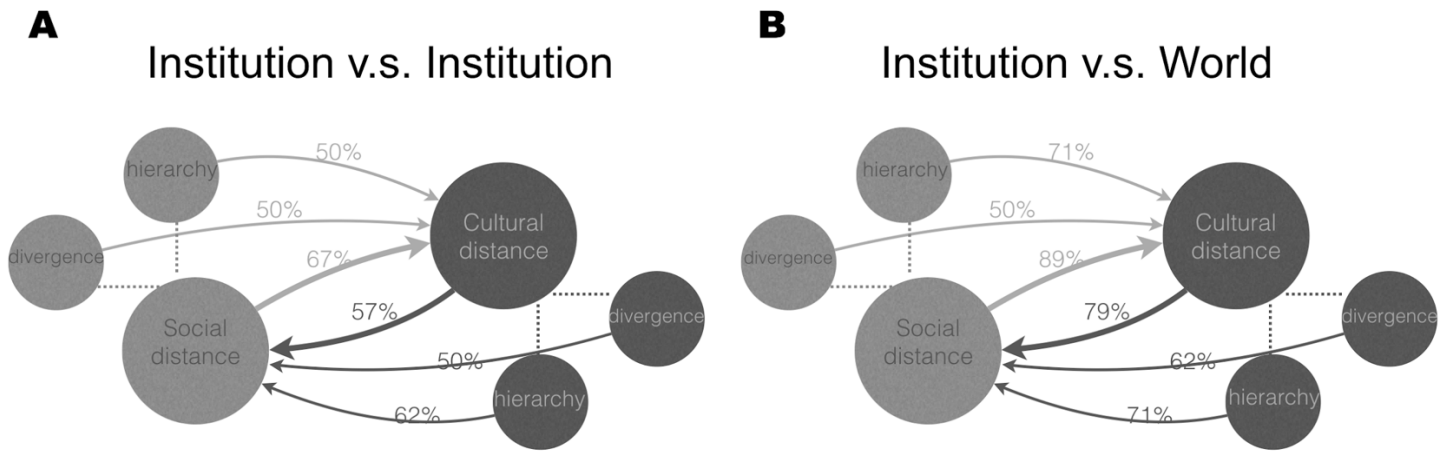

**Figure A10.** We use Granger causality regression to explore whether the social distance of institutions predicts their semantic distance in future and vice versa. In addition, as the positions of institutions and the distance between them always involves two factors, angle ("divergence") and radius ("hierarchy"), we can further investigate which factor contribute more in the prediction. We investigate the dynamics between pairwise institutions ("institution vs institution", panel A), and also aggregate the results to explore how single institutions interact with all other institutions ("institution vs world", panel B). In both panels, arrows point from the predictors to the predicted variables. The values associated with arrows were the percentages of positive coefficients in the Granger causality regression, which quantifies the predicting power in the positive direction. In general, the social distance tends to predict cultural distances slightly more than the converse, but both are consistent and significant.

Below we provide additional examples of hyperbolic embedding in the academic context. Figure A11 shows how the aggregated density distribution of PACS codes for four top universities precisely characterizes their topic focus in Physics. Figure A12 shows that all tags of questions in the online Physics Q&A community on Stack Exchange can be effectively embedded using the Poincaé disk model to reveal their hierarchy and similarity.

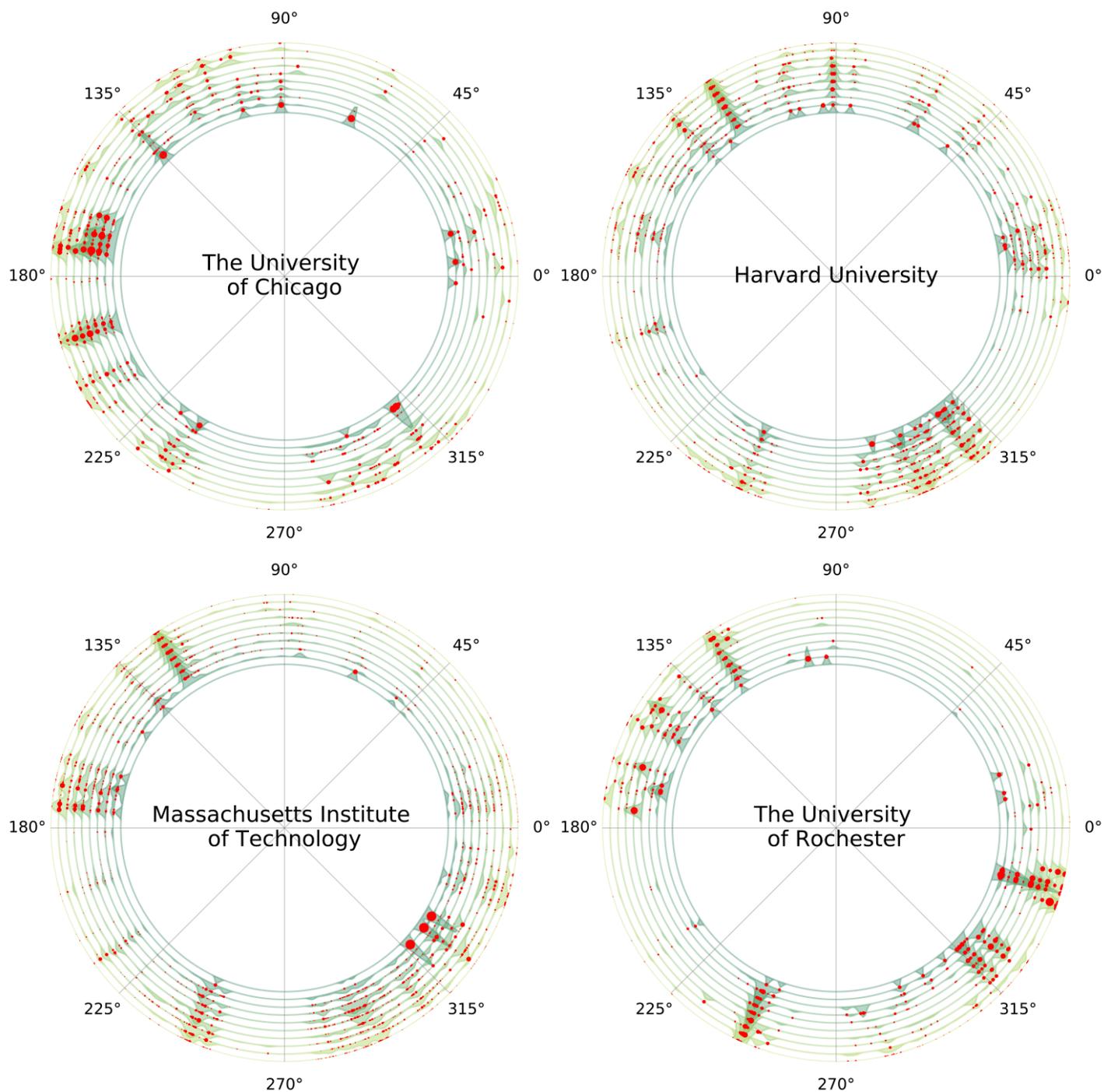

**Figure A11. Hyperbolic embedding of four universities.** The Gaussian kernel density estimation is used to visualize the concentration of the angles of PACS codes in green waves. Red dots visualize the data points of PACS codes, with a size proportional to the number of papers. Each circle represents a year from 2002 (the innermost circle) to 2011 (the outermost circle).

**Figure A12.** Hyperbolic embedding tags from 47,282 questions in Physics Stack Exchange (https://physics.stackexchange.com/).

# References

Anderson, Chris. 2006. *The Long Tail: Why the Future of Business Is Selling Less of More*. Hachette Books.
Anderson, P. W. 1972. "More Is Different." *Science*. https://doi.org/10.1126/science.177.4047.393.
Asch, Solomon E. 1956. "Studies of Independence and Conformity: I. A Minority of One against a Unanimous Majority." *Psychological Monographs: General and Applied* 70 (9): 1.
Barabasi, A. L., and R. Albert. 1999. "Emergence of Scaling in Random Networks." *Science* 286 (5439): 509–12.
Basov, Nikita, and Julia Brennecke. 2017. "Duality Beyond Dyads: Multiplex Patterning of Social Ties and Cultural Meanings." In *Structure, Content and Meaning of Organizational Networks*, 87–112. Research in the Sociology of Organizations. Emerald.
Bauer, Henry H. 1990. "Barriers Against Interdisciplinarity: Implications for Studies of Science, Technology, and Society (STS." *Science, Technology & Human Values* 15 (1): 105–19.
Becker, Joshua, Devon Brackbill, and Damon Centola. 2017. "Network Dynamics of Social Influence in the Wisdom of Crowds." *Proceedings of the National Academy of Sciences of the United States of America* 114 (26): E5070–76.
Bettencourt, Luís M. A., José Lobo, Dirk Helbing, Christian Kühnert, and Geoffrey B. West. 2007. "Growth, Innovation, Scaling, and the Pace of Life in Cities." *Proceedings of the National Academy of Sciences of the United States of America* 104 (17): 7301–6.
Biron, Lauren. 2015. "Our Flat Universe." *Symmetry: Dimensions of Particle Physics*, April 7, 2015. https://www.symmetrymagazine.org/article/april-2015/our-flat-universe?email_issue=725.
Bolukbasi, Tolga, Kai-Wei Chang, James Y. Zou, Venkatesh Saligrama, and Adam T. Kalai. 2016. "Man Is to Computer Programmer as Woman Is to Homemaker? Debiasing Word Embeddings." In *Advances in Neural Information Processing Systems*, 4349–57.
Bonola, Roberto. 1955. *Non-Euclidean Geometry: A Critical and Historical Study of Its Development*. Courier Corporation.
Boudreau, Kevin J., and Karim R. Lakhani. 2015. "'Open' Disclosure of Innovations, Incentives and Follow-on Reuse: Theory on Processes of Cumulative Innovation and a Field Experiment in Computational Biology." *Research Policy* 44 (1): 4–19.
Bourdieu, Pierre. 1984. "Distinction. Translated by Richard Nice." *Cambridge, MA: Harvard Univer*.
———. 1989. "Social Space and Symbolic Power." *Sociological Theory* 7 (1): 14–25.
Braun, Tibor, and András Schubert. 2003. "A Quantitative View on the Coming of Age of Interdisciplinarity in the Sciences 1980-1999." *Scientometrics* 58 (1): 183–89.
Breiger, Ronald L., and Kyle Puetz. 2015. "Culture and Networks." *The International Encyclopedia of the Social and Behavioral Sciences*, 557–62.
Burt, Ronald S. 2004. "Structural Holes and Good Ideas." *The American Journal of Sociology* 110 (2): 349–99.
———. 2007. *Brokerage and Closure: An Introduction to Social Capital*. OUP Oxford.
———. 2009. *Structural Holes: The Social Structure of Competition*. Harvard University Press.
Buskens, Vincent, and Arnout van de Rijt. 2008. "Dynamics of Networks If Everyone Strives for Structural Holes." *The American Journal of Sociology* 114 (2): 371–407.
Caliskan, Aylin, Joanna J. Bryson, and Arvind Narayanan. 2017. "Semantics Derived Automatically from Language Corpora Contain Human-like Biases." *Science* 356 (6334): 183–86.
Carley, Kathleen. 1986. "An Approach for Relating Social Structure to Cognitive Structure." *The Journal of Mathematical Sociology* 12 (2): 137–89.
———. 1991. "A Theory of Group Stability." *American Sociological Review* 56 (3): 331–54.
Carlino, Gerald A., Satyajit Chatterjee, and Robert M. Hunt. 2006. "Urban Density and the Rate of Invention," August. https://doi.org/10.2139/ssrn.926328.
Cetina, Karin Knorr. 2009. *Epistemic Cultures: How the Sciences Make Knowledge*. Harvard University Press.
Chamberlain, Benjamin Paul, James Clough, and Marc Peter Deisenroth. 2017. "Neural Embeddings of Graphs

in Hyperbolic Space." *arXiv [stat.ML]*. arXiv. http://arxiv.org/abs/1705.10359.

Danchev, Valentin, Andrey Rzhetsky, and James A. Evans. 2019. "Centralized Communities More Likely Generate Non-Replicable Results." *eLife*. http://arxiv.org/abs/1801.05042.

Danescu-Niculescu-Mizil, Cristian, Lillian Lee, Bo Pang, and Jon Kleinberg. 2012. "Echoes of Power: Language Effects and Power Differences in Social Interaction." In *Proceedings of the 21st International Conference on World Wide Web*, 699–708. WWW '12. New York, NY, USA: ACM.

Danescu-Niculescu-Mizil, Cristian, Robert West, Dan Jurafsky, Jure Leskovec, and Christopher Potts. 2013. "No Country for Old Members: User Lifecycle and Linguistic Change in Online Communities." In *Proceedings of the 22Nd International Conference on World Wide Web*, 307–18. WWW '13. New York, NY, USA: ACM.

Deegan, Robert D., Olgica Bakajin, Todd F. Dupont, Greb Huber, Sidney R. Nagel, and Thomas A. Witten. 1997. "Capillary Flow as the Cause of Ring Stains from Dried Liquid Drops." *Nature* 389 (October): 827.

Devlin, Jacob, Ming-Wei Chang, Kenton Lee, and Kristina Toutanova. 2018. "BERT: Pre-Training of Deep Bidirectional Transformers for Language Understanding." *arXiv [cs.CL]*. arXiv. http://arxiv.org/abs/1810.04805.

DiMaggio, Paul. 1987. "Classification in Art." *American Sociological Review*, 440–55.

Dunham, Douglas, and Others. 2009. "The Symmetry of 'Circle Limit IV' and Related Patterns." *BRIDGES*, 163–68.

Elberse, Anita. 2008. "Should You Invest in the Long Tail?" *Harvard Business Review* 86 (7/8): 88.

Erickson, Bonnie H. 1988. "The Relational Basis of Attitudes." *Social Structures: A Network Approach* 99 (121): 443–75.

Feld, Scott L. 1991. "Why Your Friends Have More Friends Than You Do." *American Journal of Sociology*. https://doi.org/10.1086/229693.

Fellbaum, Christiane. 2005. "WordNet and Wordnets." In *Encyclopedia of Language and Linguistics*, edited by Alex Barber, 2–665. Elsevier.

FIRTH, and J. R. 1957. "A Synopsis of Linguistic Theory, 1930-1955." *Studies in Linguistic Analysis*. https://ci.nii.ac.jp/naid/10020680394/.

Fleming, Lee, Santiago Mingo, and David Chen. 2007. "Collaborative Brokerage, Generative Creativity, and Creative Success." *Administrative Science Quarterly* 52 (3): 443–75.

Fourcade, Marion, Etienne Ollion, and Yann Algan. 2015. "The Superiority of Economists." *Revista de Economía Institucional* 17 (33): 13–43.

Freeman, Linton C. 1977. "A Set of Measures of Centrality Based on Betweenness." *Sociometry* 40 (1): 35–41.

Furnas, G. W., S. Deerwester, S. T. Dumais, T. K. Landauer, R. A. Harshman, L. A. Streeter, and K. E. Lochbaum. 1988. "Information Retrieval Using a Singular Value Decomposition Model of Latent Semantic Structure." In *Proceedings of the 11th Annual International ACM SIGIR Conference on Research and Development in Information Retrieval*, 465–80. SIGIR '88. New York, NY, USA: ACM.

Garg, Nikhil, Londa Schiebinger, Dan Jurafsky, and James Zou. 2018. "Word Embeddings Quantify 100 Years of Gender and Ethnic Stereotypes." *Proceedings of the National Academy of Sciences of the United States of America* 115 (16): E3635–44.

Garicano, Luis. 2000. "Hierarchies and the Organization of Knowledge in Production." *The Journal of Political Economy* 108 (5): 874–904.

Goel, Sharad, Andrei Broder, Evgeniy Gabrilovich, and Bo Pang. 2010. "Anatomy of the Long Tail: Ordinary People with Extraordinary Tastes." In *Proceedings of the Third ACM International Conference on Web Search and Data Mining*, 201–10. WSDM '10. New York, NY, USA: ACM.

Granovetter, Mark. 1974. "Finding a Job: A Study on Contacts and Careers." Cambridge, Mass.: Harvard University Press.

Granovetter, Mark S. 1973. "The Strength of Weak Ties." *The American Journal of Sociology* 78 (6): 1360–80.

Guimerà, Roger, Brian Uzzi, Jarrett Spiro, and Luís A. Nunes Amaral. 2005. "Team Assembly Mechanisms Determine Collaboration Network Structure and Team Performance." *Science* 308 (5722): 697–702.

Harari, Yuval Noah. 2014. *Sapiens: A Brief History of Humankind*. Random House.

Hong, Lu, and Scott E. Page. 2004. "Groups of Diverse Problem Solvers Can Outperform Groups of High-Ability Problem Solvers." *Proceedings of the National Academy of Sciences of the United States of*

*America* 101 (46): 16385–89.

Huntington, Samuel P. 2000. "The Clash of Civilizations?" In *Culture and Politics*, 99–118. Springer.

Hutchinson, Ian. 2011. *Monopolizing Knowledge*. Lulu.com.

Jaeger, H. M., and S. R. Nagel. 1992. "Physics of the Granular State." *Science* 255 (5051): 1523–31.

Johnson, William B., and Joram Lindenstrauss. 1984. "Extensions of Lipschitz Mappings into a Hilbert Space." *Contemporary Mathematics* 26 (189-206): 1.

Joshi, Aparna, and Hyuntak Roh. 2009. "The Role of Context in Work Team Diversity Research: A Meta-Analytic Review." *Academy of Management Journal. Academy of Management* 52 (3): 599–627.

Kandel, Denise B. 1978. "Homophily, Selection, and Socialization in Adolescent Friendships." *The American Journal of Sociology* 84 (2): 427–36.

Klemm, Konstantin, Víctor M. Eguíluz, Raúl Toral, and Maxi San Miguel. 2003. "Nonequilibrium Transitions in Complex Networks: A Model of Social Interaction." *Physical Review. E, Statistical, Nonlinear, and Soft Matter Physics* 67 (2 Pt 2): 026120.

Kozlowski, Austin C., Matt Taddy, and James A. Evans. 2018. "The Geometry of Culture: Analyzing Meaning through Word Embeddings." *arXiv [cs.CL]*. arXiv. http://arxiv.org/abs/1803.09288.

———. 2019. "The Geometry of Culture: Analyzing the Meanings of Class through Word Embeddings." *American Sociological Review* 84 (5): 905–49.

Krackhardt, David, and Martin Kilduff. 1990. "Friendship Patterns and Culture: The Control of Organizational Diversity." *American Anthropologist* 92 (1): 142–54.

Krioukov, Dmitri, Maksim Kitsak, Robert S. Sinkovits, David Rideout, David Meyer, and Marián Boguñá. 2012. "Network Cosmology." *Scientific Reports* 2 (November): 793.

Krioukov, Dmitri, Fragkiskos Papadopoulos, Maksim Kitsak, Amin Vahdat, and Marián Boguñá. 2010. "Hyperbolic Geometry of Complex Networks." *Physical Review. E, Statistical, Nonlinear, and Soft Matter Physics* 82 (3 Pt 2): 036106.

Krohn, Marvin D. 1986. "The Web of Conformity: A Network Approach to the Explanation of Delinquent Behavior." *Social Problems* 33 (6): s81–93.

Kuhn, Thomas S. 1970. *The Structure of Scientific Revolutions*.

Latour, Bruno. 1987. *Science in Action: How to Follow Scientists and Engineers Through Society*. Harvard University Press.

Lazarsfeld, Paul F., Robert K. Merton, and Others. 1954. "Friendship as a Social Process: A Substantive and Methodological Analysis." *Freedom and Control in Modern Society* 18 (1): 18–66.

Lee, Monica, and John Levi Martin. 2018. "Doorway to the Dharma of Duality." *Poetics* 68 (June): 18–30.

Lewis, Kevin, and Jason Kaufman. 2018. "The Conversion of Cultural Tastes into Social Network Ties." *The American Journal of Sociology* 123 (6): 1684–1742.

Leydesdorff, Loet, and Inga A. Ivanova. 2014. "Mutual Redundancies in Interhuman Communication Systems: Steps toward a Calculus of Processing Meaning." *Journal of the Association for Information Science and Technology* 65 (2): 386–99.

Lizardo, Omar. 2014. "Omnivorousness as the Bridging of Cultural Holes: A Measurement Strategy." *Theory and Society* 43 (3): 395–419.

Lorenz, Jan, Heiko Rauhut, Frank Schweitzer, and Dirk Helbing. 2011. "How Social Influence Can Undermine the Wisdom of Crowd Effect." *Proceedings of the National Academy of Sciences* 108 (22): 9020–25.

Malone, Elizabeth. 2005. "Linguistics and Languages; A Special Report: Endangered Languages." National Science Foundation. 2005. https://www.nsf.gov/news/special_reports/linguistics/endangered.jsp.

Mannix, Elizabeth, and Margaret A. Neale. 2005. "What Differences Make a Difference? The Promise and Reality of Diverse Teams in Organizations." *Psychological Science in the Public Interest: A Journal of the American Psychological Society* 6 (2): 31–55.

March, James G. 1991. "Exploration and Exploitation in Organizational Learning." *Organization Science* 2 (1): 71–87.

Mark, Noah. 1998. "Birds of a Feather Sing Together." *Social Forces; a Scientific Medium of Social Study and Interpretation* 77 (2): 453–85.

Martin, John Levi, and Monica Lee. 2018. "A Formal Approach to Meaning." *Poetics* 68 (June): 10–17.

McFarland, Daniel A., Dan Jurafsky, and Craig Rawlings. 2013. "Making the Connection: Social Bonding in Courtship Situations." *The American Journal of Sociology* 118 (6): 1596–1649.

McPherson, Miller, Lynn Smith-Lovin, and James M. Cook. 2001. "Birds of a Feather: Homophily in Social Networks." *Annual Review of Sociology* 27 (1): 415–44.

Mikolov, Tomas, Ilya Sutskever, Kai Chen, Greg S. Corrado, and Jeff Dean. 2013. "Distributed Representations of Words and Phrases and Their Compositionality." In *Advances in Neural Information Processing Systems 26*, edited by C. J. C. Burges, L. Bottou, M. Welling, Z. Ghahramani, and K. Q. Weinberger, 3111–19. Curran Associates, Inc.

Mohr, John W. 2000. "Introduction: Structures, Institutions, and Cultural Analysis." *Poetics* 27 (2): 57–68.

Montgomery, Georgina M., and Mark A. Largent. 2015. *A Companion to the History of American Science*. John Wiley & Sons.

Nickel, Maximillian, and Douwe Kiela. 2017. "Poincaré Embeddings for Learning Hierarchical Representations." In *Advances in Neural Information Processing Systems 30*, edited by I. Guyon, U. V. Luxburg, S. Bengio, H. Wallach, R. Fergus, S. Vishwanathan, and R. Garnett, 6338–47. Curran Associates, Inc.

Nielsen, Michael. 2012. *Reinventing Discovery: The New Era of Networked Science*. Princeton University Press.

Osgood, Charles Egerton, George J. Suci, and Percy H. Tannenbaum. 1964. *The Measurement of Meaning*. University of Illinois Press.

Pachucki, Mark A., and Ronald L. Breiger. 2010. "Cultural Holes: Beyond Relationality in Social Networks and Culture." *Annual Review of Sociology* 36 (1): 205–24.

Page, Scott E. 2008. *The Difference: How the Power of Diversity Creates Better Groups, Firms, Schools, and Societies - New Edition*. Princeton University Press.

Palla, Gergely, Albert-László Barabási, and Tamás Vicsek. 2007. "Quantifying Social Group Evolution." *Nature* 446 (April): 664.

Papadopoulos, Fragkiskos, Maksim Kitsak, M. Ángeles Serrano, Marián Boguñá, and Dmitri Krioukov. 2012. "Popularity versus Similarity in Growing Networks." *Nature* 489 (7417): 537–40.

Papadopoulos, Fragkiskos, Constantinos Psomas, and Dmitri Krioukov. 2015. "Network Mapping by Replaying Hyperbolic Growth." *IEEE/ACM Transactions on Networking* 23 (1): 198–211.

Pennington, Jeffrey, Richard Socher, and Christopher Manning. 2014. "Glove: Global Vectors for Word Representation." In *Proceedings of the 2014 Conference on Empirical Methods in Natural Language Processing (EMNLP)*, 1532–43.

Perozzi, Bryan, Rami Al-Rfou, and Steven Skiena. 2014. "DeepWalk: Online Learning of Social Representations." In *Proceedings of the 20th ACM SIGKDD International Conference on Knowledge Discovery and Data Mining*, 701–10. KDD '14. New York, NY, USA: ACM.

Poon, WCK. 2002. "The Physics of a Model Colloid–polymer Mixture." *Journal of Physics. Condensed Matter: An Institute of Physics Journal*. http://iopscience.iop.org/article/10.1088/0953-8984/14/33/201/meta.

Prigogine, I. 1947. "Le Rendement Thermodynamique de La Thermodiffusion." *Physica*. https://doi.org/10.1016/0031-8914(47)90003-7.

Prigogine, Ilya, and Raymond Defay. 1944. *Thermodynamique Chimique Conformément Aux Méthodes de Gibbs et De Donder*. Vol. 2. Desoer.

Regier, Terry. 2008. "Whorf Was Half Right." *PsycEXTRA Dataset*. https://doi.org/10.1037/e527312012-204.

Rogers, Everett M. 2010. *Diffusion of Innovations, 4th Edition*. Simon and Schuster.

Rogers, Everett M., and F. Floyd Shoemaker. 1971. "Communication of Innovations; A Cross-Cultural Approach." https://eric.ed.gov/?id=ED065999.

Roth, Camille, and Jean-Philippe Cointet. 2010. "Social and Semantic Coevolution in Knowledge Networks." *Social Networks* 32 (1): 16–29.

Saint-Charles, Johanne, and Pierre Mongeau. 2009. "Different Relationships for Coping with Ambiguity and Uncertainty in Organizations." *Social Networks* 31 (1): 33–39.

Salganik, Matthew J., and Duncan J. Watts. 2008. "Leading the Herd Astray: An Experimental Study of Self-Fulfilling Prophecies in an Artificial Cultural Market." *Social Psychology Quarterly*. https://doi.org/10.1177/019027250807100404.

Shannon, Claude, and Warren Weaver. 1963. *The Mathematical Theory of Communication*. University of Illinois Press.
Shi, Feng, Misha Teplitskiy, Eamon Duede, and James Evans. 2017. "The Wisdom of Polarized Crowds." *arXiv [cs.SI]*. arXiv. http://arxiv.org/abs/1712.06414.
Shuicheng Yan, Dong Xu, Benyu Zhang, and Hong-Jiang Zhang. 2005. "Graph Embedding: A General Framework for Dimensionality Reduction." In *2005 IEEE Computer Society Conference on Computer Vision and Pattern Recognition (CVPR'05)*, 2:830–37 vol. 2.
Simon, Herbert A. 1955. "On a Class of Skew Distribution Functions." *Biometrika* 42 (3/4): 425–40.
Strang, David, and Michael W. Macy. 2001. "In Search of Excellence: Fads, Success Stories, and Adaptive Emulation." *American Journal of Sociology*. https://doi.org/10.1086/323039.
Surowiecki, James. 2004. "The Wisdom of Crowds: Why the Many Are Smarter than the Few and How Collective Wisdom Shapes Business." *Economies, Societies and Nations* 296.
Uzzi, Brian, Satyam Mukherjee, Michael Stringer, and Ben Jones. 2013. "Atypical Combinations and Scientific Impact." *Science* 342 (6157): 468–72.
Uzzi, Brian, and Jarrett Spiro. 2005. "Collaboration and Creativity: The Small World Problem." *The American Journal of Sociology* 111 (2): 447–504.
Vaan, Mathijs de, Balazs Vedres, and David Stark. 2015. "Game Changer: The Topology of Creativity1." *AJS; American Journal of Sociology* 120 (4): 1–51.
Valente, Thomas W. 1996. "Network Models of the Diffusion of Innovations." *Computational & Mathematical Organization Theory* 2 (2): 163–64.
Vedres, Balázs, and David Stark. 2010. "Structural Folds: Generative Disruption in Overlapping Groups." *The American Journal of Sociology* 115 (4): 1150–90.
Wasserman, Stanley, and Katherine Faust. 1994. *Social Network Analysis: Methods and Applications*. Cambridge University Press.
Wasserman, Stanley, and Philippa Pattison. 1996. "Logit Models and Logistic Regressions for Social Networks: I. An Introduction to Markov Graphs Andp." *Psychometrika* 61 (3): 401–25.
Watts, D. J., and S. H. Strogatz. 1998. "Collective Dynamics of 'Small-World' Networks." *Nature* 393 (6684): 440–42.
Whorf, Benjamin Lee, and Stuart Chase. 1956. *Language, Thought and Reality, Selected Writings of Benjamin Lee Whorf. Edited... by John B. Carroll. Foreword by Stuart Chase*. Mass.
Wittgenstein, Ludwig, Gertrude Elizabeth Margaret Anscombe, and Ludwig Wittgenstein. 1953. *Philosophical Investigations... Translated by GEM Anscombe.(Philosophische Untersuchungen.) Eng. & Ger*. Oxford.
Woolley, Anita Williams, Christopher F. Chabris, Alex Pentland, Nada Hashmi, and Thomas W. Malone. 2010. "Evidence for a Collective Intelligence Factor in the Performance of Human Groups." *Science* 330 (6004): 686–88.
Wuchty, Stefan, Benjamin F. Jones, and Brian Uzzi. 2007. "The Increasing Dominance of Teams in Production of Knowledge." *Science* 316 (5827): 1036–39.
Wu, Lingfei, Dashun Wang, and James A. Evans. 2017. "Large Teams Have Developed Science and Technology; Small Teams Have Disrupted It," September. https://papers.ssrn.com/sol3/papers.cfm?abstract_id=3034125.
Young, Hugh D., and Roger A. Freedman. 2013. *University Physics with Modern Physics Technology Update: International Edition*. Pearson Educacion.
Zuckerman, Ezra W., and John T. Jost. 2001. "What Makes You Think You're so Popular? Self-Evaluation Maintenance and the Subjective Side of the 'Friendship Paradox.'" *Social Psychology Quarterly*. https://doi.org/10.2307/3090112.